\documentclass[longauth]{aa}
\usepackage{graphicx}
\usepackage{txfonts}
\usepackage{xcolor}
\usepackage{hyperref}
\hypersetup{colorlinks, linkcolor=blue, citecolor=blue, urlcolor=blue}

\newcommand{\te}{t_{\rm E}}

\def\e{{\rm E}}

\newcommand{\bdv}[1]{\mbox{\boldmath$#1$}}

\def\au{{\rm au}} 
 
\def\kms{{\rm km}\,{\rm s}^{-1}}
\def\masyr{{\rm mas}\,{\rm yr}^{-1}}
\def\kpc{{\rm kpc}}
\def\mas{{\rm mas}}

\def\muas{\mu{\rm as}}

\def\pc{{\rm pc}}

\def\max{{\rm max}}

\def\rel{{\rm rel}}

\def\eff{{\rm eff}}
\def\rot{{\rm rot}}
\def\lsr{{\rm lsr}}
\def\hel{{\rm hel}}

\def\e{{\rm E}}
\def\bpi{{\bdv\pi}}
\def\bmu{{\bdv\mu}}

\def\bgamma{{\bdv\gamma}}

\def\bv{{\bf v}}

\definecolor{brown}{rgb}{0.59, 0.29, 0.0}
\definecolor{darkgreen}{rgb}{0.0, 0.42, 0.24}
\definecolor{darkblue}{rgb}{0.01, 0.31, 0.59}
\definecolor{darkblue}{rgb}{0.0, 0.25, 0.42}
\definecolor{blue}{rgb}{0.0,0.0,1.0}
\definecolor{green}{rgb}{0.0,1.0,0.0}

\def\eqalign#1{\null\,\vcenter{\openup\jot
        \ialign{\strut\hfil$\displaystyle{##}$&$
        \displaystyle{{}##}$\hfil \crcr#1\crcr}}\,}

\begin{document}

\title{Systematic KMTNet Planetary Anomaly Search. V.  Complete Sample of 2018 Prime-Field}

\author{
     Andrew Gould\inst{1,2} 
\and Cheongho Han\inst{3}
\and Weicheng Zang\inst{4}
\and Hongjing Yang\inst{4} 
\and Kyu-Ha Hwang\inst{5}
\and Andrzej Udalski\inst{6}
\and Ian~A.~Bond\inst{7}
\\
(Leading authors)\\
     Michael D. Albrow\inst{8} 
\and Sun-Ju Chung\inst{5} 
\and Youn Kil Jung\inst{5} 
\and Yoon-Hyun Ryu\inst{5} 
\and In-Gu Shin\inst{3} 
\and Yossi Shvartzvald{\inst9} 
\and Jennifer~C.~Yee\inst{10} 
\and Sang-Mok Cha\inst{5,11} 
\and Dong-Jin Kim\inst{5}
\and Hyoun-Woo Kim\inst{5} 
\and Seung-Lee Kim\inst{5,12} 
\and Chung-Uk Lee\inst{5} 
\and Dong-Joo Lee\inst{5}
\and Yongseok Lee\inst{5,11} 
\and Byeong-Gon Park\inst{5,12} 
\and Richard W. Pogge\inst{2}
\\
(The KMTNet Collaboration)\\
     Przemek Mr{\'o}z\inst{6}
\and Micha{\l} K. Szyma{\'n}ski\inst{6}
\and Jan Skowron\inst{6}
\and Radek Poleski\inst{6}
\and Igor Soszy{\'n}ski\inst{6}  
\and Pawe{\l} Pietrukowicz{\inst6}
\and Szymon Koz{\l}owski\inst{6}
\and Krzysztof Ulaczyk\inst{13}
\and Krzysztof A. Rybicki\inst{6}
\and Patryk Iwanek\inst{6}
\and Marcin Wrona\inst{6}
\\
(The OGLE Collaboration)\\
     Fumio Abe\inst{14} 
\and Richard Barry\inst{15} 
\and David P. Bennett\inst{15,16} 
\and Aparna Bhattacharya\inst{15,16} 
\and Hirosame Fujii\inst{14}   
\and Akihiko~Fukui\inst{17,18} 
\and Yuki Hirao\inst{19}
\and Stela Ishitani Silva\inst{20,15,14}
\and Rintaro Kirikawa\inst{19}
\and Iona Kondo\inst{19} 
\and Naoki Koshimoto\inst{21} 
\and Yutaka Matsubara\inst{14} 
\and Sho Matsumoto\inst{19} 
\and Shota Miyazaki\inst{19} 
\and Yasushi Muraki\inst{14} 
\and Arisa Okamura\inst{19}   
\and Greg Olmschenk\inst{15}
\and Cl{\'e}ment Ranc\inst{22} 
\and Nicholas J. Rattenbury\inst{23} 
\and Yuki Satoh\inst{19}
\and Takahiro Sumi\inst{19}
\and Daisuke Suzuki\inst{19} 
\and Taiga Toda\inst{19}
\and Paul~J.~Tristram{24} 
\and Aikaterini Vandorou\inst{15,16}
\and Hibiki Yama\inst{19}
\\
(The MOA Collaboration) \\
     Charles Beichman\inst{25}
\and Geoffry Bryden\inst{26}
\and Sebastiano Calchi Novati\inst{25}
\and B. Scott Gaudi\inst{2}
\and Calen B. Henderson\inst{25}
\and Matthew~T.~Penny\inst{27}
\and Savannah Jacklin\inst{28}
\and Keivan G. Stassun\inst{28}
\\
(The UKIRT Microlensing Team) \\
}

\institute{
     Max-Planck-Institute for Astronomy, K\"{o}nigstuhl 17, 69117 Heidelberg, Germany                                                                       
\and Department of Astronomy, Ohio State University, 140 W.  18th Ave., Columbus, OH 43210, USA                                                             
\and Department of Physics, Chungbuk National University, Cheongju 28644, Republic of Korea, corresponding author, \email{cheongho@astroph.chungbuk.ac.kr}  
\and  Department of Astronomy, Tsinghua University, Beijing 100084, China                                                                                   
\and Korea Astronomy and Space Science Institute, Daejon 34055, Republic of Korea                                                                           
\and Astronomical Observatory, University of Warsaw, Al.~Ujazdowskie~4, 00-478~Warszawa, Poland                                                             
\and Institute of Natural and Mathematical Science, Massey University, Auckland 0745, New Zealand                                                           
\and University of Canterbury, Department of Physics and Astronomy, Private Bag 4800, Christchurch 8020, New Zealand                                        
\and Department of Particle Physics and Astrophysics, Weizmann Institute of Science, Rehovot 76100, Israel                                                  
\and  Center for Astrophysics $|$ Harvard \& Smithsonian, 60 Garden St., Cambridge, MA 02138, USA                                                           
\and School of Space Research, Kyung Hee University, Yongin, Kyeonggi 17104, Republic of Korea                                                              
\and Korea University of Science and Technology, Korea, (UST), 217 Gajeong-ro, Yuseong-gu, Daejeon, 34113, Republic of Korea                                
\and Department of Physics, University of Warwick, Gibbet Hill Road, Coventry, CV4~7AL,~UK                                                                  
\and Institute for Space-Earth Environmental Research, Nagoya University, Nagoya 464-8601, Japan                                                            
\and Code 667, NASA Goddard Space Flight Center, Greenbelt, MD 20771, USA                                                                                   
\and Department of Astronomy, University of Maryland, College Park, MD 20742, USA                                                                           
\and Department of Earth and Planetary Science, Graduate School of Science, The University of Tokyo, 7-3-1 Hongo, Bunkyo-ku, Tokyo 113-0033, Japan          
\and Instituto de Astrof{\'i}sica de Canarias, V{\'i}a L{\'a}ctea s/n, E-38205 La Laguna, Tenerife, Spain                                                   
\and Department of Earth and Space Science, Graduate School of Science, Osaka University, Toyonaka, Osaka 560-0043, Japan                                   
\and Department of Physics, The Catholic University of America, Washington, DC 20064, USA                                                                   
\and Department of Astronomy, Graduate School of Science, The University of Tokyo, 7-3-1 Hongo, Bunkyo-ku, Tokyo 113-0033, Japan                            
\and Sorbonne Universit\'e, CNRS, UMR 7095, Institut d'Astrophysique de Paris, 98 bis bd Arago, 75014 Paris, France                                         
\and Department of Physics, University of Auckland, Private Bag 92019, Auckland, New Zealand                                                                
\and University of Canterbury Mt. John Observatory, P.O. Box 56, Lake Tekapo 8770, New Zealand                                                              
\and IPAC, Mail Code 100-22, Caltech, 1200 E. California Blvd., Pasadena, CA 91125                                                                          
\and Jet  Propulsion  Laboratory,  California  Institute  of  Technology,  4800  Oak  Grove  Drive,  Pasadena,  CA 91109, USA                               
\and Department of Physics and Astronomy, Louisiana State University, Baton Rouge, LA 70803 USA                                                             
\and Department of Physics \& Astronomy,Vanderbilt University, Nashville, TN 37235, USA                                                                     
}
\date{Received ; accepted}

\abstract
{
We complete the analysis of all 2018 prime-field microlensing planets
identified by the KMTNet AnomalyFinder. 
Among the 10 previously unpublished 
events with clear planetary solutions, 8 are either unambiguously
planetary or are very likely to be planetary in nature:
OGLE-2018-BLG-1126, 
KMT-2018-BLG-2004, 
OGLE-2018-BLG-1647, 
OGLE-2018-BLG-1367, 
OGLE-2018-BLG-1544, 
OGLE-2018-BLG-0932, 
OGLE-2018-BLG-1212,  and
KMT-2018-BLG-2718.
Combined with the 4 previously published new AnomalyFinder events
and 12 previously published (or in preparation) planets that were discovered
by eye, this makes a total of 24 2018 prime-field planets
discovered or recovered by AnomalyFinder.  Together with a paper in
preparation on 2018 sub-prime planets, this work lays the basis for the
first statistical analysis of the planet mass-ratio function based on
planets identified in KMTNet data.
By systematically applying the
heuristic analysis of \citet{kb190253} to each event, we identify the
small modification in their formalism that is needed to unify the
so-called close/wide and inner/outer degeneracies, as conjectured by
\citet{ob190960}.
}


\maketitle

\section{{Introduction}
\label{sec:intro}}

From its inception, and even conception, the Korea Microlensing Telescope
Network (KMTNet, \citealt{kmtnet}) had as its major aim the construction
and analysis of a large-scale statistical sample of microlensing planets.
Nevertheless, during its first five years of full operations (2016-2020),
the overwhelming focus was on the detection and analysis of individual
events of high scientific interest.  In part, this focus reflected the
new possibilities opened by KMTNet's continuous wide field coverage
from three continents.  For example, KMTNet played a major or decisive
role in the detections of all three of the planets with mass ratios
$q< 3\times 10^{-5}$ that were known by 2020 \citep{kb180029,ob190960,kb200414}

During this period, substantial work was carried out that would ultimately
lay the basis for large-scale statistical studies.  This included
the development of a tiered observing strategy covering $97\,{\rm deg}^2$
of the Galactic bulge (Figure 12 of \citealt{eventfinder}), as well
as robust methods of identifying of order 3000 microlensing events per year
using the EventFinder and AlertFinder systems \citep{eventfinder,alertfinder}.

However, a number of practical, technical, and scientific challenges
impeded the inauguration of large-scale statistical studies.  At the
most basic level, the online photometry remained of mixed quality until
2019.  This did not prevent high-precision analysis of individual events
because, from the beginning, KMTNet had a tender-loving care (TLC) system
of data re-reduction
based on pySIS \citep{albrow09}, which returned high-quality photometry on 
an event-by-event basis.  However, it did mean that automated planet-searches
of the KMTNet database would have yielded difficult-to-interpret results.  
In 2019, a new end-of-season pipeline was put into
place that produced good-quality photometry for the great majority of
events.  This enabled the first KMTNet statistical study, a search for
free-floating planet (FFP) candidates in the 2019 database \citep{kb192073}.
The same pipeline was gradually applied to the three previous seasons,
but this labor-intensive work was only completed in November 2021.

Making use of these improved databases, \citet{ob191053} developed a 
new AnomalyFinder algorithm that was adapted to the characteristics of
KMTNet, i.e., combining unprecedented quantities of microlensing data from
three sites operating under very different conditions.  The key innovation
was to fit for ``anomalies'' in the residuals rather than for planets
in the original light curves, which permitted a reduction of the
search from three to two dimensions and also vastly simplified the modeling.
This dimensional reduction is adapted from the KMTNet EventFinder algorithm
\citep{gould96,eventfinder}, and like EventFinder, it results in many
false positives for each true anomaly, which must then be rejected by human
review.  However, in contrast to EventFinder, which annually results in 
${\cal O}(5\times 10^5)$ false positives on ${\cal O}(5\times 10^8)$ 
catalog stars, the AnomalyFinder yields ${\cal O}(1\times 10^4)$ false 
positives on ${\cal O}(3\times 10^3)$ microlensing events.  That is,
while the specific false-positive rate is larger by 3.5 orders of magnitude,
the total number of false positives is smaller by a factor 50, making
human review much more tractable.  In particular, it is quite
feasible for several people to independently conduct this review as
a cross check.

The specific false-positive rate is larger because the search is much
more aggressive, i.e., attempting to discover all planetary anomalies
down to a very low threshold.  In particular, for AnomalyFinder, the
operator may be shown dozens of potential anomalies, whereas for 
EventFinder only the highest-$\chi^2$ candidate event is shown.  
Stated otherwise,
the search can be much more aggressive because the number of light curves
has been reduced from $5\times 10^8$ to $3\times 10^3$, i.e., by $10^5$.

Another practical obstacle was the large human effort required for
TLC reductions, which often took of order one day of work for each event.
Again, this is not a major problem if one is publishing of order a dozen
events per year.  However, a statistical analysis requires not only
the accurate parameter characterization of all ``interesting'' planets,
but of all planets, and more dauntingly, all anomalous (or potentially
anomalous) events that might plausibly be planetary.  We estimate that
this will be of order 200 TLC reductions for 2016-2019.  Motivated
by this challenging situation, H.\ Yang et al. (2022, in preparation) developed
a quasi-automated TLC system that reduces the average reduction time to about
one hour.

Our immediate goal is to prepare a complete sample of AnomalyFinder
events from 2018 that can be compared to planet detection efficiency 
calculator (Jung et al.\ 2022, in preparation).  This will be the first
step toward the analysis of the 2016-2019 sample. 

In the present paper, we complete the prime-field sample, i.e.,
all planets found in KMTNet fields with nominal cadences 
$\Gamma\geq 2\,{\rm hr}^{-1}$, specifically BLG01, BLG02, BLG03, 
BLG41, BLG42, and BLG43.  The updated AnomalyFinder2.0 \citep{af2}
identified a total of 114 anomalous events (from an underlying sample of 843
prime-field events), which it classified as
``planet'' (23), ``planet/binary'' (16), ``binary/planet'' (18),
``binary'' (53), and ``finite source'' (4).
Among the 53 in the ``binary'' classification, 14 were judged by
eye to be unambiguously non-planetary in nature.  Among the 23 in the
``planet'' classification, 13 were either previously published (11)
or in preparation (2).  Among the 16 in the ``planet/binary'' classification,
one (KMT-2018-BLG-0748) was a previously published planet, 
and among the 18 in the ``binary/planet'' classification 
one (OGLE-2018-BLG-1544) was previously known to have a planetary solution.
See Table~11 of \citet{kb190253}.

\begin{table*}[t]
\small
\caption{Event Names, Cadences, Alerts, and Locations}
\begin{tabular}{lcccccc}
\hline\hline
\multicolumn{1}{c}{Name}                      & 
\multicolumn{1}{c}{$\Gamma\,({\rm hr}^{-1})$} &
\multicolumn{1}{c}{Alert Date}                &
\multicolumn{1}{c}{RA$_{\rm J2000}$}          &
\multicolumn{1}{c}{Dec$_{\rm J2000}$}         &
\multicolumn{1}{c}{$l$}                       &
\multicolumn{1}{c}{$b$}                       \\
\hline   
OGLE-2018-BLG-1126     & 0.3  & 22 Jun 2018 & 17:53:25.41  & $-31$:43:28.99  & $-1.53$  & $-2.88$  \\
KMT-2018-BLG-2064      & 4.0                                                                       \\
\hline
KMT-2018-BLG-2004      & 4.0  & Post Season & 17:53:42.58  & $-30$:20:25.26  & $-0.30$  & $-2.23$  \\ 
\hline
OGLE-2018-BLG-1647     & 0.3  & 7 Sep 2018  & 17:55:50.97  & $-31$:49:01.20  & $-1.35$  & $-3.37$  \\
KMT-2018-BLG-2060      & 4.0                                                                       \\
\hline
OGLE-2018-BLG-1367     & 0.6  & 6 Aug 2018  & 17:59:01.35  & $-29$:10:06.10  & $+1.29$  & $-2.64$  \\
MOA-2018-BLG-320       & 4.0                                                                       \\
KMT-2018-BLG-0914      & 2.0                                                                       \\
\hline
OGLE-2018-BLG-1544     & 1.0  & 17 Aug 2018 & 17:56:30.92  & $-30$:24:10.30  & $-0.05$  & $-2.78$  \\
KMT-2018-BLG-0787      & 4.0                                                                       \\
\hline
OGLE-2018-BLG-0932     & 1.0  & 28 May 2018 & 17:53:24.29  & $-29$:01:18.50  & $+0.81$  & $-1.50$  \\
KMT-2018-BLG-2087      & 2.0                                                                       \\
MOA-2018-BLG-163$^a$   & 4.0                                                                       \\
\hline
OGLE-2018-BLG-1212     & 1.0  & 9 Jul 2018  & 18:04:18.63  & $-28$:11:38.70  & $+2.72$  & $-3.17$  \\
MOA-2018-BLG-365       & 4.0                                                                       \\
KMT-2018-BLG-2299      & 4.0                                                                       \\
\hline
KMT-2018-BLG-2718      & 4.0  & Post Season & 17:52:56.08  & $-30$:13:06.28  & $-0.28$ & $-2.02$   \\ 
\hline
KMT-2018-BLG-2164      & 4.0  & Post Season & 17:57:22.83  & $-30$:09:57.20  & $+0.25$ & $-2.83$   \\ 
\hline
OGLE-2018-BLG-1554     & 0.2  & 17 Aug 2018 & 17:57:59.28  & $-31$:26:05.89  & $-0.79$  & $-3.57$  \\ 
MOA-2018-BLG-329       & 1.3                                                                       \\ 
KMT-2018-BLG-0809      & 4.0                                                                       \\
\hline
\end{tabular}
\label{tab:names}
\end{table*}

All of  the remaining 85 events were fitted using online data, i.e.,
pipeline photometry.  Of these, four new planets have already been published,
including three by
\citet{kb190253} in a study of low-$q$ planets, and one by
\citet{ob180383} as part of a study of wide-orbit planets.
Of the remaining 81, 56 were found to have $q>0.06$, one 
(OGLE-2018-BLG-2014) was determined to
be in the range $(0.05<q<0.06)$, and 24 required TLC reductions, either
because they were potentially planetary, $q_{\rm online}<0.05$, or because
the light curve could not be reliably characterized without TLC reductions.
Of these 24, the 7 that have planetary solutions are analyzed here.
Note that the 28 events that required TLC (24 analyzed here, and
four previously published planets), were distributed among the five
classification categories (planet, planet/binary, binary/planet, binary, finite
source) as (9,11,4,3,1) of which (8,1,0,0,1) ultimately proved to have
unambiguous planetary solutions and (1,1,0,0,0) ultimately proved to have
planetary solutions that were ambiguous.  
This shows that great majority of events that
ultimately prove to have planetary solutions are classified at the
first stage as ``planets'' and that the great majority of events so classified
prove to be planetary.
We also analyze 3 of the 4 such events that were listed as ``in preparation''
in Table~11 of \citep{kb190253} (namely, 
OGLE-2018-BLG-0932,
OGLE-2018-BLG-1554, and
OGLE-2018-BLG-1647)
), for a total of 10 events with  planetary
solutions\footnote{From detailed analysis, the remaining event, 
OGLE-2018-BLG-0100/KMT-2018-BLG-2296, is known to be planetary in
nature but with competitive solutions that differ in $q$ by a factor 100,
so that it is not suitable for mass-ratio function studies.}.  
These 10 include 8 that are clearly  or very likely planetary 
in nature ($q<0.03$) and 2 others that have an ambiguous nature. 
One additional event (OGLE-2018-BLG-0856/KMT-2018-BLG-2392) was found 
to have a solution with $0.03<q<0.05$ based on TLC data.  We follow a policy
of noting such events but excluding them from the sample.

In sum, based on previous analyses and the current work,
the prime-field sample has a total 26 planets or possible planets,
of which 23 have unambiguous mass-ratio determinations, making them potentially
suitable for a statistical analysis.  Note that these must still be
vetted for various effects, for which we provide detailed guidance in the
text.  The 3 others are clearly 
unsuitable because they are subject to multiple interpretations in $q$.

Note that the fraction of events that were subjected to AnomalyFinder analysis
that were initially classified as planet and/or binary ($110/843=13\%$)
and those finally determined to be planetary or possibly planetary
($26/843=3.1\%$) are both very similar the corresponding ratios in
the first high-cadence 24/7 microlensing survey that was carried out
by \citet{shvartzvald16}.  They found $26/244 = 13\%$ anomalous events and
$9/244 = 3.8\%$ planets, i.e., both identical within Poisson errors.

For 2018, AnomalyFinder2.0 has already been run on the 21 KMTNet fields
with lower cadence, $\Gamma < 2\,{\rm hr}^{-1}$, covering $84\,{\rm deg}^2$
and yielding a total of 173 anomalous events, 
which are distributed among the five classifications as (17, 4, 19, 126, 7).
These include nine published planets and three in preparation.
However, among the nine published planets, three have ambiguous or larger-error
mass-ratio measurements and so, are not suitable for studying the
planet-host mass-ratio function.  Therefore, we may expect that after
lower-cadence AnomalyFinder output is fully studied, 
there will be a total of 30--35 planets in the 2018 sample.

This will be comparable in size to the largest previous
study \citep{suzuki16}, which included 22 planets from six years
of the Microlensing Observations in Astrophysics II (MOA-II) survey,
augmented by 8 planets found in two earlier survey/followup studies
\citep{gould10,cassan12}.
However, the AnomalyFinder sample 
will subsequently be expanded to cover at least four years,
as we continue to publish all planets 2016-2019.

\section{{Observations}
\label{sec:obs}}

As described in Section~\ref{sec:intro}, all of the planetary 
(or potentially planetary) events that are presented
in this paper were initially identified by applying the AnomalyFinder2.0
algorithm \citep{af2} to the 843 events that were originally found by the
KMTNet EventFinder and AlertFinder systems in the prime fields during 2018.
As described by \citet{kb190253}, when available, we use data 
from independent alerts from the Optical Gravitational Lensing Experiment (OGLE)
and MOA to vet the anomalies for systematics (otherwise, we study these
anomalies at the image level).  We also include OGLE and MOA data
in the analysis of the events.  These were taken using the OGLE 1.3m telescope
with $1.4\,{\rm deg}^2$ field of view at Las Campanas Observatory in Chile,
and the MOA 1.8m telescope
with $2.2\,{\rm deg}^2$ field of view at Mt.\ John Observatory in New Zealand.
The OGLE and MOA data analyzed here are in the $I$ band and a broad, customized,
$R$-$I$ filter, respectively.

Table~\ref{tab:names} gives basic
observational information about each event.  Column~1 gives the
event names in the order of discovery (if discovered by multiple teams),
which enables cross identification.  However, in most of the
rest of the paper, we refer to events only by the name given by the group
who made the first discovery.  The nominal cadences are given in column 2,
and column 3 shows the first discovery date.  The remaining four columns
show the event coordinates in the equatorial and galactic systems.
Events with OGLE names were originally discovered by the OGLE Early
Warning System \citep{ews1,ews2}.  

To the best of our knowledge, there were no ground-based follow-up observations
of any of these events.  One event, OGLE-2018-BLG-0932, lies in the field of the
UKIRT Microlensing Survey \citep{ukirt17}, and we make use of these
data to determine its source color.  This survey employs a 3.8m telescope
in Hawaii, with an effective field of view of $0.2\,{\rm deg}^2$, to observe 
in the $H$ and $K$ bands.
OGLE-2018-BLG-0932 was also observed
by the {\it Spitzer} space telescope, but the analysis of the resulting
data is beyond the scope of the present work and will be presented elsewhere.

The KMT, OGLE, and MOA data were reduced using difference image analysis 
\citep{tomaney96,alard98},
as implemented by each group, i.e., \citet{albrow09},
\citet{wozniak2000}, and \citet{bond01}, respectively.  The UKIRT data were
reduced using the CASU multi-aperture
photometry pipeline producing 2MASS $H$- and $K$-band calibrated
magnitudes \citep{irwin04,hodgkin09}.

\section{{Light Curve Analysis}
\label{sec:anal}}

\subsection{{Preamble}
\label{sec:anal-preamble}}

We begin by describing the light-curve analysis methods and notations
that are common to all events.
All of the events in this paper appear, to a first approximation as
simple 1L1S light curves, which can be described by three \citet{pac86}
parameters, $(t_0,u_0,t_\e)$, i.e., the time of lens-source closest
approach, the impact parameter in units of $\theta_\e$ and the Einstein
timescale,
\begin{equation}
t_\e = {\theta_\e\over\mu_\rel}; \ \ \ 
\theta_\e = \sqrt{\kappa M\pi_\rel}; \ \ \ 
\kappa\equiv {4\,G\over c^2\,\au} \simeq 8.14\,{\mas\over M_\odot},
\label{eqn:tedef}
\end{equation}
where $M$ is the lens mass, $\pi_\rel$ and $\bmu_\rel$ are the
lens-source relative parallax and proper-motion, respectively,
and $\mu_\rel \equiv |\bmu_\rel|$.  Here $n$L$m$S means $n$ lenses
and $m$ sources.   In addition, to these three non-linear parameters,
two flux parameters, $(f_S,f_B)$, are required for each observatory,
representing the source flux and the blended flux that does not
participate in the event.  Note, however, that these are linear
parameters, which can be determined by regression after
the model is specified by the non-linear parameters.

We then search for ``static'' 2L1S solutions, which require four
additional parameters $(s,q,\alpha,\rho)$, i.e., the planet-host separation
in units of $\theta_\e$, the planet-host mass ratio, the angle of the
source trajectory relative to the binary axis, and the angular source
size normalized to $\theta_\e$, i.e., $\rho=\theta_*/\theta_\e$.

We conduct this search in two phases.  In the first phase, we search
on a 2-dimensional (2-D) grid.  For each $(s,q)$ pair, we construct
a magnification map following \citet{mb07400}.  We then conduct a downhill
search using the Monte Carlo Markov chain (MCMC) technique.  We seed this
search with the 1L1S solution for the \citet{pac86} parameters, $(t_0,u_0,t_\e)$.
We use the approach of \citet{gaudi02} to find the seed for $\rho$.
For $\alpha$ we seed at a grid of values around the unit circle.
This procedure yields a $\chi^2$ map on the $(s,q)$ plane, which
we use to identify one or several local minima.

In the second phase, we refine the best solution at each local minimum
by allowing all seven parameters to vary in the MCMC.

In this analysis, we often make use of a modified version of the
 heuristic analysis introduced by \citet{kb190253}.
If a brief anomaly at $t_{\rm anom}$
is assumed to be generated by the source crossing the planet-host axis,
then \citet{kb190253} suggested analytic estimates for $(s,\alpha)$ of
\begin{equation}
s = s^\dagger \pm \Delta s; \ \ \ 
s^\dagger_\pm = {\sqrt{4 + u_{\rm anom}^2}\pm u_{\rm anom}\over 2}; \ \ \ 
\tan\alpha ={u_0\over\tau_{\rm anom}},
\label{eqn:heuristic-old}
\end{equation}
where $u_{\rm anom}^2= \tau_{\rm anom}^2 + u_0^2$, 
$\tau_{\rm anom} = (t_{\rm anom}-t_0)/t_\e$, and where $\Delta s$ (i.e.,
half the difference between the two solutions) generally
cannot be determined from by-eye inspection.  In the great majority of
cases, $s^\dagger_+>1$ corresponds to anomalous bumps and
$s^\dagger_-<1$ corresponds to anomalous dips.   This formalism was designed
to reflect the ``inner/outer'' degeneracy \citep{gaudi97}
whereby the source passes the
planetary caustic(s) on the side closer to (or farther from) the
central caustic.  However, following the work of \citet{ob180677} and
\citet{ob190960}, it was already recognized to have somewhat wider application.

In the course of the present investigation, in which we systematically
applied Equation~(\ref{eqn:heuristic-old}) to all 10 events, we encountered
OGLE-2018-BLG-1647, which proved to be the ``Rosetta Stone'' that 
unified the ``inner/outer'' degeneracy for planetary caustics
\citep{gaudi97} with the ``close/wide'' degeneracy for central 
caustics \citep{griest98}, as conjectured by \citet{ob190960}.  
For this event, the formula for $s_+^\dagger$ in 
Equation~(\ref{eqn:heuristic-old}) proved to be a better approximation
to the geometric mean of the two empirically derived solutions, $s_\pm$,
i.e., $s^\dagger = \sqrt{s_+ s_-}$, rather than the arithmetic mean,
i.e., $s^\dagger = (s_+ + s_-)/2$.

This fact immediately led to several realizations.  First, this reformulation
did not contradict any of the four cases examined by \citet{kb190253},
nor the many other cases examined in the current work, because for these
$\Delta\ln s \equiv (1/2)\ln(s_+/s_-)$ was always small,
$\Delta\ln s \ll 1$.  In this limit, 
for which Equation~(\ref{eqn:heuristic-old}) worked quite well, 
the arithmetic and geometric means
differ by only $\sim(\Delta\ln s)^2/2$, which is generally too small
to notice.  Second, the mathematical representation of this reformulation,
\begin{equation}
s_\pm = s^\dagger\exp(\pm \Delta \ln s),
\label{eqn:heuristic}
\end{equation}
is equivalent to the usual expression for the ``close/wide'' degeneracy,
$s_- = 1/s_+$, provided that $s^\dagger\rightarrow 1$.  Moreover, because
\citet{griest98} derived this relation in the limit of central 
caustics, i.e., high-magnification events for which
$u_{\rm anom}\ll 1$, the limit $s^\dagger\rightarrow 1$ does indeed apply to
this case.  Third, what made OGLE-2018-BLG-1647 a ``Rosetta Stone''
is that the geometric mean of Equation~(\ref{eqn:heuristic}) applied,
even though $s^\dagger\not=1$ (contrary to the ``close/wide'' limit).
Fourth, the several historical examples that inspired \citet{ob190960}
to suggest unification were all ``inner/outer'' degeneracies in which
one of the two solutions had the source passing between the central
and planetary caustics, while the other had it passing outside the
planetary wing of a resonant caustic.  That is, one solution
appeared more closely related to the ``inner/outer'' degeneracy and the
other to the ``close/wide'' degeneracy\footnote{Prior to the work of
\citet{ob180677} and \citet{ob190960}, the inner/outer degeneracy was
conceived more narrowly as having the source pass on opposite sides
of detached planetary caustic(s).  To our knowledge, there had been only
two recognized cases of this degeneracy, i.e., 
OGLE-2016-1067 with minor-image caustics \citep{ob161067} and
OGLE-2017-0173 with a major-image caustic \citep{ob170173}.
}.  The pair of solutions were dubbed
``inner/outer'' primarily because both solutions had the same logarithmic
sign, $(\ln s_+)(\ln s_-)>0$.  This had already indicated a continuous
degeneracy to \citet{ob190960}.  However, in the course of this
(and other) work, we noted additional cases with similar topologies,
but for which $(\ln s_+)(\ln s_-)<0$ (as in the ``close/wide'' limit),
but for which Equation~(\ref{eqn:heuristic}) remained a better
approximation than the $s_- = 1/s_+$ prediction of \citet{griest98}.
We regarded this as further evidence for a continuum of $(s_-,s_+)$ degeneracies
from   inner/outer ($s^\dagger < 1$, minor-image caustics),
through close/wide ($s^\dagger \simeq 1$, central and resonant caustics),
to     outer/inner ($s^\dagger > 1$, major-image caustics).

Subsequently, \citet{kb211391} have provided uniform notation for this
formalism in their Equations (2)--(7).  We follow their conventions here.
In particular $s_\pm^\dagger$ (with ``$\pm$'' subscript)
denotes the theoretical prediction of Equation~(\ref{eqn:heuristic-old}),
while $s^\dagger$ (without subscript) denotes the geometric mean of the
two empirical solutions, whose offset is characterized by $\Delta\ln s$,
\begin{equation}
s^\dagger\equiv \sqrt{s_+s_-},
\qquad
\Delta\ln s\equiv {1\over 2}\ln {s_+\over s_-}.
\label{eqn:iocw}
\end{equation}

\citet{kb190253} also introduced an estimate of the mass ratio $q$
for dip-type anomalies, which is ultimately based on the theoretical
analysis by \citet{han06}:
\begin{equation}
q = \biggl({\Delta t_{\rm dip}\over 4\, t_\e}\biggr)^2
{s^\dagger\over |u_0|}|\sin^3\alpha|,
\label{eqn:qeval}
\end{equation}
where $\Delta t_{\rm dip}$ is the full duration of the dip.  \citet{kb211391}
noted that this expression can be rewritten in terms of
``direct observables'':
\begin{equation}
t_q \equiv q t_\e = {1\over 16}{(\Delta t_{\rm dip})^2\over t_\eff}
\biggl(1 + {(\delta t_{\rm anom})^2\over t_\eff^2}\biggr)^{-3/2}s^\dagger,
\label{eqn:qeval2}
\end{equation}
where they pointed out that
$\Delta t_{\rm dip}$ and $\delta t_{\rm anom}$ can be read directly
off the light curve, while $t_\eff\equiv u_0t_\e\simeq {\rm FWHM}/\sqrt{12}$ 
for even moderately high magnification events, $A_\max \ga 5$.  
Indeed, \citet{mb11293} had already pointed
out that $t_q=q t_\e$ is also an invariant for high-magnification events.

In some cases, we investigate whether the microlens parallax vector
\citep{gould92,gould00,gould04}
\begin{equation}
\bpi_\e\equiv {\pi_\rel\over \theta_\e}\,{\bmu_\rel\over\mu_\rel}
\label{eqn:piedef}
\end{equation}
can be constrained by the data.  Note that if this quantity can be measured,
then by combining Equations~(\ref{eqn:tedef}) and (\ref{eqn:piedef}) one
can infer the lens and mass and distance,
\begin{equation}
M = {\theta_\e\over\kappa\pi_\e}; \qquad
D_L = {\au\over \theta_\e\pi_\e + \pi_S},
\label{eqn:mpirel}
\end{equation}
where $\pi_S$ is the parallax of the source, which usually is
approximately known.  However, even if $\bpi_\e$ cannot be measured
(e.g., it is consistent with zero at $1\,\sigma$), it can significantly
constrain $(M,D_L)$ after imposing priors from a Galactic model, provided
that the error ellipse on $\bpi_\e$ is sufficiently small, at least
in one dimension (see the Appendix in \citealt{ob150479}).

To model the parallax effects due to Earth's orbital motion, we add
two parameters $(\pi_{\e,N},\pi_{\e,E})$, which are the components of
$\bpi_\e$ in equatorial coordinates.  Because these effects can be mimicked
by those due to lens orbital motion \citep{mb09387,ob09020},
we always add (at least initially)
two parameters $\bgamma =[(ds/dt)/s,d\alpha/dt]$, where
$s\bgamma$ are the first derivatives of projected lens orbital motion at $t_0$,
i.e., parallel and perpendicular to the projected separation of the
planet at that time, respectively.
In order to eliminate unphysical solutions, we impose the constraint
on the ratio of the transverse kinetic to potential energy 
\citep{eb2k5,ob05071b},
\begin{equation}
\beta \equiv \bigg|{\rm KE\over PE}\bigg|
= {\kappa M_\odot {\rm yr}^2\over 8\pi^2}\,{\pi_\e\over\theta_\e}\gamma^2
\biggr({s\over \pi_\e + \pi_S/\theta_\e}\biggr)^3 < 0.8 .
\label{eqn:betadef}
\end{equation}
Note that while orbits are only unbound if $\beta>1$, we impose a
slightly stronger constraint because it is extremely rare for planets
to be in such high-eccentricity orbits and observed at the right orientation
and epoch to yield $\beta>0.8$.

It often happens that $\bgamma$ is neither significantly constrained
nor significantly correlated with $\bpi_\e$.  In these cases, we suppress
these two degrees of freedom.

Very frequently, including several cases in this paper, the parallax
contours in the $\bpi_\e$ plane take the form of elongated ellipses
\citep{gmb94}
with the orientation angle of short axis, $\psi$, being approximately 
aligned with the projected position angle of the Sun, $\psi_\odot$, at the
peak of the event, $t_0$.  That is, $\psi\simeq\psi_\odot$.  This is because,
for events with $t_\e\ll 1\,$yr, Earth's acceleration is approximately
constant, under which condition lens-source motion along the direction
of acceleration gives rise to much more pronounced effects than does
the transverse motion \citep{smp03}.
When this occurs, it can be substantially more informative to characterize
$\bpi_\e=(\pi_{\e,\parallel},\pi_{\e,\perp})$ in terms of these two components.
For example, unless $\psi$ is closely aligned with one of the cardinal
directions, $\sigma(\pi_{\e,\parallel})$ can be much smaller than either
$\sigma(\pi_{\e,N})$ or $\sigma(\pi_{\e,E})$.  For reference, we note that
the (Gaussian) likelihood associated with the parallax measurement can
be expressed as,
\begin{equation}
L(\bpi_\e) = {\exp\biggl[-\sum_{i=1}^2\sum_{j=1}^2
(\bpi_\e- \bpi_{\e,0})_i b_{i,j} (\bpi_\e- \bpi_{\e,0})_j\biggr]
\over 2\pi \sigma(\pi_{\e,\parallel})           \sigma(\pi_{\e,\perp})}
\label{eqn:pielike}
\end{equation} 
where $\bpi_{\e,0}$ is the best fit,
$b\equiv c^{-1}$, $c$ is the covariance matrix, and where we have
written the determinant of this matrix explicitly in terms of its eigenvectors
in order to make contact with future applications.

As pointed out by \citet{gaudi98}, 1L2S events can mimic 2L1S events,
particularly if there are no sharp caustic-crossing features in the light curve.
If $\Delta\chi^2 = \chi^2({\rm 1L2S}) - \chi^2({\rm 2L1S})$ is strongly 
negative, then we conclude that the event is 1L2S, and we eliminate it
from consideration.  If we test for 1L2S and find that $\Delta\chi^2$ 
is strongly positive, we remark that such solutions are ruled out.
If 1L2S and 2L1S have either competitive or roughly
comparable $\chi^2$ we report both solutions.
The former class of events are ambiguous in nature and
cannot be included in planetary catalogs, nor
certainly in mass-ratio function studies.  However, we report such events
because it may be possible in the future to resolve the degeneracy for some
of them using auxiliary data.

We carry out 1L2S modeling by adding at least 3 parameters 
$(t_{0,2},u_{0,2},q_F)$ to the 3 \citet{pac86} parameters.
These are the time of closest approach and impact parameter of the
second source and the ratio of the second to the first source flux
in the $I$-band \citep{1L2S}.  
If either lens-source approach can be interpreted
as exhibiting finite source effects, then we must add one or two further
parameters, i.e., $\rho_1$ and/or $\rho_2$.  And, if the two sources
are projected closely enough on the sky, one must also consider
source orbital motion (e.g., \citealt{ob151459}).

\begin{figure}[t]
\includegraphics[width=\columnwidth]{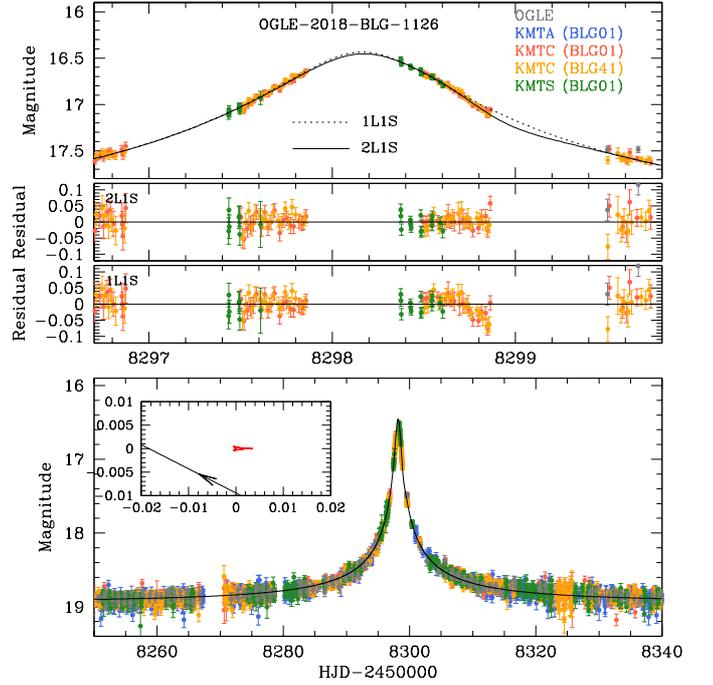}
\caption{Light curve and model for OGLE-2018-BLG-1126.
The anomaly is a dip that is centered at 8298.7, which
is detected at $\Delta\chi^2 =\chi^2({\rm 1L1S})- \chi^2({\rm 2L1S})=69$.
As in all 10 light-curve figures in this paper, we show the full
light curve and anomaly region in separate panels, we show the
caustic topologies in one or more insets, we show residual
panels for indicated models, and we color the data points by observatory,
as indicated in the legend.}
\label{fig:2064lc}
\end{figure}

\subsection{{OGLE-2018-BLG-1126}
\label{sec:anal-ob181126}}

The KMTC data exhibit a systematic
decline relative to the 1L1S model centered on 8298.7.
See Figure~\ref{fig:2064lc}.
The formal significance of this deviation is modest:
$\Delta\chi^2 = \chi^2({\rm 1L1S}) - \chi^2({\rm 2L1S})= 69$.  Moreover,
because the coverage of the anomaly is incomplete, one must be concerned
that this deviation is due to some systematic effect.  The main potential
cause of such an effect would be the Moon, which was full when it
passed through the bulge (about $11^\circ$ north of the event)
roughly 36 hours before the anomaly.  There is a well-known mechanism
for the Moon to induce a spurious excess (though not deficit) in the 
tabulated flux, which generates many false alerts of short timescale 
events by the EventFinder \citep{eventfinder}: the higher background
pushes a bright star above the pixel well depth, causing charge to bleed
into a column and so pollute the photometry of fainter stars that
are downstream in the same column.  
These bleeds are often invisible on normal displays
of the original images because the stretch is generally too weak to
detect them.  However, they are easily visible on difference images,
for which the stretch can be made much stronger.  We carefully examine
the difference images throughout the night and find no such signatures.
Another possibility is that the Moon caused excess flux on the
previous night when it resulted in much higher background (13000 versus 4000),
thus affecting the overall light-curve model, thereby giving the
appearance of an anomaly on the following night.  However, we
see no evidence for bleeds on the previous night.  
Thus we conclude that the anomaly is real.

Adopting \citet{pac86} parameters 
$(t_0,u_0,t_\e)=(8298.17,0.0083,53.3\,{\rm day})$ 
and light curve features $(t_{\rm dip},\Delta t_{\rm dip})=(8298.8,1.2{\rm day})$,
the heuristic approach outlined in Section~\ref{sec:anal-preamble} yields
$\tau_{\rm anom}=+0.0118$, $\alpha=35^\circ$, $s^\dagger_-=0.993$, and
$q=7\times 10^{-4}$.

\begin{table}[t]
\small
\caption{Light Curve Parameters for OGLE-2018-BLG-1126}
\begin{tabular*}{\columnwidth}{@{\extracolsep{\fill}}lcccc}
\hline\hline
\multicolumn{1}{c}{Parameter}    &
\multicolumn{1}{c}{Close}        &
\multicolumn{1}{c}{Wide }        \\
\hline
$t_0 -8290$                   & $ 8.1661\pm 0.0036$ & $ 8.1679\pm 0.0034$\\
$u_0\ (10^{-2})$              & $  0.830\pm  0.058$ & $  0.824\pm  0.053$\\
$t_\e$ (days)                 & $  53.26\pm   3.40$ & $  53.33\pm   3.15$\\
$s$                           & $  0.852\pm  0.040$ & $  1.154\pm  0.052$\\
$q\ (10^{-3})$                & $  0.082\pm  0.048$ & $  0.059\pm  0.040$\\
$\langle\log q\rangle$        & $  -4.13\pm   0.28$ & $  -4.26\pm   0.29$\\
$\alpha$ (rad)                & $  0.496\pm  0.038$ & $  0.528\pm  0.036$\\
$\rho^a\ (10^{-3})$           &                     &                    \\
$I_{\rm S}$                   & $  21.58\pm   0.07$ & $  21.58\pm   0.07$\\
\hline
\end{tabular*}
\tablefoot{a: No useful limit could be placed upon $\rho$.}
\label{tab:ob1126parms}
\end{table}

The grid search returns two local minima.  After refining these
as described in Section~\ref{sec:anal-preamble}, we find that they
generally agree with heuristic prediction.  See Table~\ref{tab:ob1126parms}.  
The main discrepancy is
in $\alpha$ ($29^\circ$ versus $35^\circ$), which is mainly due to the
difficulty of judging the center of dip from the incomplete light curve.
Of particular note is the striking agreement of 
$s^\dagger\equiv\sqrt{s_+ s_-} = 0.992$
(compared to the prediction $s^\dagger_- = 0.993$).  Thus, although
this degeneracy would normally be considered as a classic example
of the ``close/wide degeneracy'' for central and resonant caustics
because $s_{\rm close}s_{\rm wide}\simeq 1$,
the prediction of the $s^\dagger$ formalism (derived in the limit of
planetary caustics) is actually 10 times more accurate\footnote{Note
that in the geometric-mean formalism of Equation~(\ref{eqn:heuristic}),
$s^\dagger = \sqrt{s_+s_-}<1$, which conforms to the minor-image caustic
morphology of the light curve.  However, this is not true of the
arithmetic-mean prediction $s^\dagger = (s_++s_-)/2>1$,  This was a 
significant puzzle for us when we initially analyzed this event, but
was resolved after analyzing the ``Rosetta Stone'' event OGLE-2018-BLG-1647.
See Section~\ref{sec:anal-preamble}}.
Note that there is essentially no constraint on $\rho$ for this planet.

Due to the faintness of the source, we do not attempt a parallax analysis.

While we have concluded that the planet is real, it may not be
suitable for mass-ratio function studies.  From Table~\ref{tab:ob1126parms},
we see that
the $1\,\sigma$ error in $\log q$ is 0.28 dex, which corresponds to
a factor of $\sim 1.9$.  The goal of the present paper is not to
impose a boundary for this parameter, but rather to present a comprehensive
account of all planets that meet much broader criteria in order to
provide a basis for such choices in future analyses of the mass-ratio function.
However, we remark that it is at least questionable whether this planet
will enter such studies.

We note that although this planet meets the $q<2\times 10^{-4}$
selection criterion of \citet{kb190253}, it was not included in
their sample.  This is because it was detected by AnomalyFinder2.0
\citep{af2}, but not AnomalyFinder1.0 \citep{ob191053}, which was the basis
of the \citet{kb190253} study.

\begin{figure}[t]
\includegraphics[width=\columnwidth]{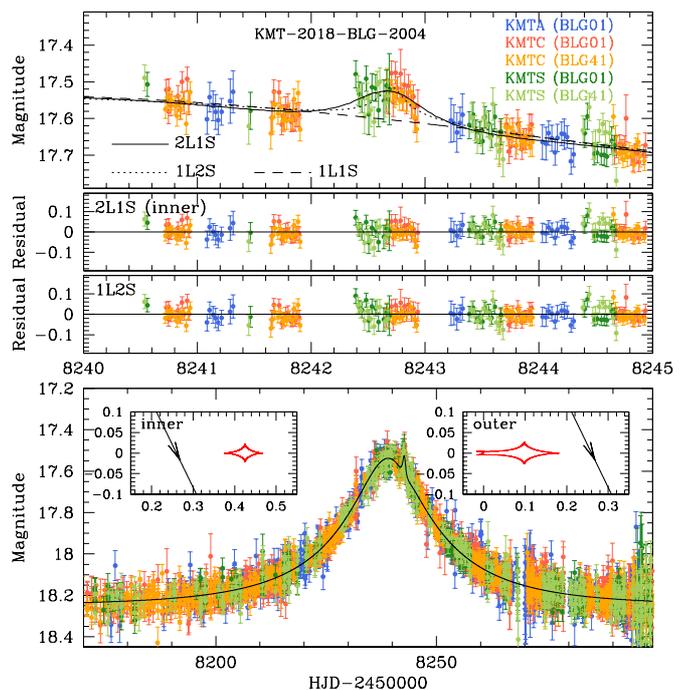}
\caption{Light curve and model for KMT-2018-BLG-2004.  The anomaly is
a bump centered at 8242.7.  The planetary interpretation is favored
over the binary-source model by
$\Delta\chi^2 =\chi^2({\rm 1L2S})- \chi^2({\rm 2L1S})=14.8$.
By including $V$-band data, this becomes $\Delta\chi^2 = 15.1$.
}
\label{fig:2004lc}
\end{figure}

\subsection{{KMT-2018-BLG-2004}
\label{sec:anal-kb182004}}

The anomaly in Figure~\ref{fig:2004lc} 
consists of a short bump, which is traced by both
KMTS and KMTC data, centered on $t_{\rm anom}=8242.7$, when the
Moon was about $10^\circ$ north of the event.
While in this case, the Moon was 4 days past full (so the background
at passage was 5000, compared to 13000 for OGLE-2018-BLG-1126),
it is far more plausible that the Moon would cause a bump in the light
curve, rather than a dip.  Indeed, given that the bump is continuous
across two observatories separated by 8000 km, it is difficult to
conceive of any other source of systematics.  However, we again
carefully examine the subtracted images and find no evidence of
bleeding columns.  Hence, we again conclude that the anomaly is
due to microlensing.

\begin{table}[t]
\small
\caption{Light Curve Parameters for KMT-2018-BLG-2004}
\begin{tabular*}{\columnwidth}{@{\extracolsep{\fill}}lcccc}
\hline\hline
\multicolumn{1}{c}{Parameter}    &
\multicolumn{1}{c}{Inner}        &
\multicolumn{1}{c}{Outer}        &
\multicolumn{1}{c}{1L2S }        \\
\hline
$\chi^2/$dof                  & $ 7308.88/ 7454$   & $ 7307.71/ 7454$   & $ 7322.51/ 7454$    \\
$t_0 -8230$                   & $  9.166\pm  0.030$& $  9.156\pm  0.033$& $  9.063\pm  0.035$ \\
$t_{0,2} -8230$               &                    &                    & $ 12.708\pm  0.032$ \\
$u_0\ (10^{-2})$              & $  23.38\pm   0.84$& $  23.19\pm   0.91$& $  23.24\pm   0.94$ \\
$u_{0,2}\ (10^{-2})$          &                    &                    & $  -0.35\pm   0.43$ \\
$t_\e$ (days)                 & $  31.26\pm   0.79$& $  31.44\pm   0.88$& $  31.73\pm   0.90$ \\
$s$                           & $  1.230\pm  0.020$& $  1.062\pm  0.017$&                     \\
$q\ (10^{-3})$                & $   0.41\pm   0.12$& $   0.37\pm   0.10$&                     \\
$\langle\log q\rangle$        & $ -3.39\pm  0.12$  & $ -3.43\pm  0.11$  &                     \\
$\alpha$ (rad)                & $  4.265\pm  0.008$& $  4.262\pm  0.009$&                     \\
$\rho\ (10^{-3})$             & $< 21$             & $< 24$             &                     \\
 $\rho_2\ (10^{-3})$          &                    &                    &                     \\
$q_f$                         &                    &                    & $0.00219\pm0.00051$ \\
$I_{\rm S}$                   & $  19.46\pm   0.05$& $  19.48\pm   0.05$& $  19.49\pm   0.05$ \\
\hline
\end{tabular*}
\label{tab:kb2004parms}
\end{table}

Using the above $t_{\rm anom}$, combined with the 1L1S parameters
$(t_0,u_0,t_\e) = (8239.17,0.23,31\,{\rm day})$, the 
heuristic formalism (see Equation~(\ref{eqn:heuristic})) predicts
$s_+^\dagger= 1.14$ and $\alpha = 244^\circ$.  The grid search returns only
two solutions, which after refinement agree quite well with these
predictions. See Table~\ref{tab:kb2004parms}.  In particular, 
$s^\dagger=\sqrt{s_{\rm inner}s_{\rm outer}} = 1.14$.
The anomaly is detected at $\chi^2({\rm 1L1S}) - \chi^2({\rm 2L1S})=167$.

Given that the anomaly is a featureless bump, it is essential to 
check whether it can be explained by a binary source (1L2S) model.
From Table~\ref{tab:kb2004parms},
we see that such models are disfavored by $\Delta\chi^2=14.8$,
which is substantial, though not overwhelming, evidence in favor of 2L1S.

In the 1L2S model, the best fit value of the flux ratio is 
$q_F=2.2\times 10^{-3}$, corresponding to a magnitude difference of
$\Delta I= -2.5\log(q_F) = 6.6$ magnitudes.  We will show in 
Section~\ref{sec:cmd-kb182004} that the source lies about 3.6 mag below
the clump.  Hence, the putative source companion would have an absolute
magnitude of $M_{I,\rm comp}\sim 10$.  Such stars are common, so the 1L2S 
solution cannot be regarded as implausible on these grounds.

The 1L2S model makes the definite prediction that the ``bump'' should
be basically invisible in the $V$ band.  That is, the source companion
should have $(V-I)_{\rm comp,0}\sim 3.3$ whereas (as we show in 
Section~\ref{sec:cmd-kb182004}), $(V-I)_{\rm S,0} \sim 0.7$.  Thus,
the relative amplitude of the bump should be $10^{0.4(3.3-0.7)}= 11$ times
smaller in $V$ than $I$.  This implies that the $V$-band light curve should 
follow the $I$-band light curve for 2L1S but should follow the 1L1S curve for
1L2S.  See \citet{ob151459}.
Unfortunately, the $V$ data are not good enough to test this
prediction.  Of the four potential data sets, 
(KMTC \& KMTS)$\times$(BLG01 \& BLG41), only KMTS BLG01 provides useful
information.  This has only one $V$-band point over the bump.  The point
lies almost exactly on the 2L1S curve.  However, it is only $0.5\,\sigma$
from the 1L1S curve, due to the relative large $V$-band error bars.

\begin{figure}[t]
\includegraphics[width=\columnwidth]{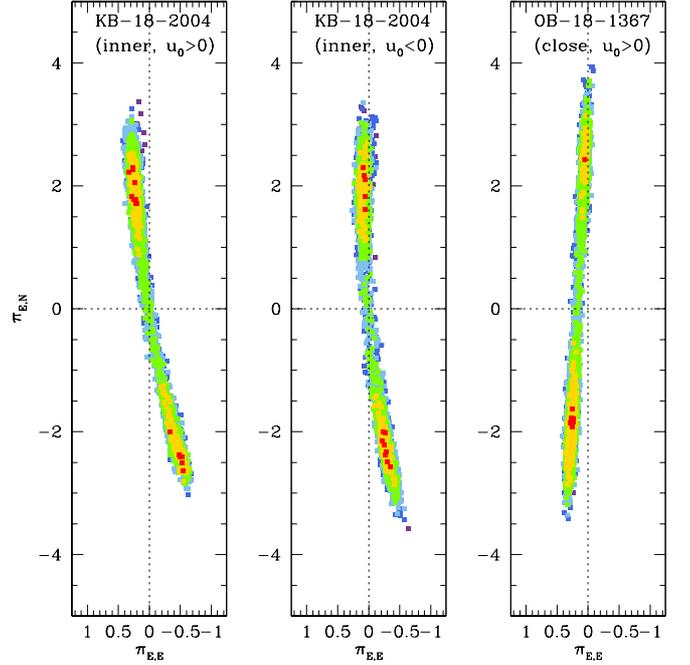}
\caption{Parallax contours for KMT-2018-BLG-2004 and OGLE-2018-BLG-1367.
For both events, these contours have very large axis ratios that are
characteristic of so-called 1-D parallax measurements.  We argue in the
text that only the short-axis information in these contours is reliable
and reduce them to truly 1-D constraints.  See 
Equations~(\ref{eqn:pielike}) and (\ref{eqn:bmatrix}) and
Sections~\ref{sec:anal-kb182004} and \ref{sec:anal-ob181367}.
}
\label{fig:parallax}
\end{figure}

Thus, the only strong argument against the 1L2S solution is that
$\Delta\chi^2=14.8$.  (If we incorporate the $V$-band test, just mentioned,
this becomes $\Delta\chi^2=15.1$).
We consider that the planet solution is strongly preferred, but we cannot
rule out the binary-source solution unconditionally.

The event is moderately long and has good photometry, so we attempt
to fit it for parallax.  Figure~\ref{fig:parallax} shows the parallax contours
for two of the four cases, namely the ``inner'' solution with $u_0>0$,
and $u_0<0$.

The parallax fit reveals interesting
information.  The basic form is of a so-called 1-dimensional (1-D) parallax
measurement, which occurs because Earth's
acceleration toward the projected position of the Sun ($\psi_\odot = 96.7^\circ$
north though east) is roughly constant over the relatively short duration
of the event.  See Section~\ref{sec:anal-preamble}.
Formally the error ellipses have an access ratio of $\sim 12$.  
The two ``lobes''
toward the north and south imply that the measurement is subject to
the so-called jerk-parallax degeneracy \citep{gould04,mb03037}.  While these are
striking to the eye, in part because of their large values, $\pi_\e\sim 2$,
they are favored by only $\Delta\chi^2\sim 4$, which would have marginal
statistical significance even if the errors could be treated as Gaussian.
That is, even in this case, their weight would be overwhelmed by the
Galactic priors in a Bayesian analysis, which heavily disfavors such large
parallax values.  Moreover, in addition to having
larger statistical errors along the long axis of the ellipse, the result
is also more subject to systematic errors because the information
is coming primarily from the wings of the light curve \citep{smp03,gould04}.

The actual information in these contours comes from their small width,
not their best-fit values. In principle, if these narrow contours
were displaced from the origin, as in the first microlensing planet
with such features, OGLE-2005-BLG-071 \citep{ob05071b}, then they would
be strong evidence for a minimum value $\pi_\e\geq\pi_{\e,\parallel}$, 
even if the exact
value was not determined.  However, in the present case, the contours
pass through the origin, so the result has less discriminatory value.

Nevertheless, we proceed to extract the essence of the parallax
information, while suppressing possible systematic effects, by retaining
the short-axis information $\sigma(\pi_{\e,_\parallel})$, while setting 
$\sigma(\pi_{\e,\perp})\rightarrow\infty$,
and using the fact that the contours pass through the origin.  Noting
that the contours ``bend'' at the origin, we adopt for the four
cases (${\rm sgn}(u_0)=\pm;{\rm sgn}(\pi_{\e,N})=\pm)$),
$({\rm sgn}(u_0),{\rm sgn}(\pi_{\e,N}),\sigma(\pi_{\e,\parallel}),\psi)) =
 (+,+,0.0453, 94.29^\circ)$,
$(+,-,0.0482,104.87^\circ)$,
$(-,+,0.0509, 89.17^\circ)$, and
$(-,-,0.0446, 99.76^\circ)$.
Then, when applying Equation~(\ref{eqn:pielike}) in 
Section~\ref{sec:phys-kb182004}, we evaluate the inverse covariance matrix
$b$ in the (north,east) equatorial system as
\begin{equation}
b(N,E) ={1\over [\sigma(\pi_{\e,\parallel})]^2}
\begin{pmatrix}
\cos^2\psi       & \sin\psi\cos\psi \\
\sin\psi\cos\psi & \sin^2\psi       \\
\end{pmatrix}
\label{eqn:bmatrix}
\end{equation}
and we set $\bpi_{\e,0}=0$.  Because this is a 1-D constraint (albeit on
a 2-D space), we substitute 
$2\pi \sigma(\pi_{\e,\parallel})\pi \sigma(\pi_{\e,\perp})
\rightarrow \sqrt{2\pi} \sigma(\pi_{\e,\parallel})$.
Note that, by construction, $b$ is a degenerate matrix.

\begin{figure}[t]
\includegraphics[width=\columnwidth]{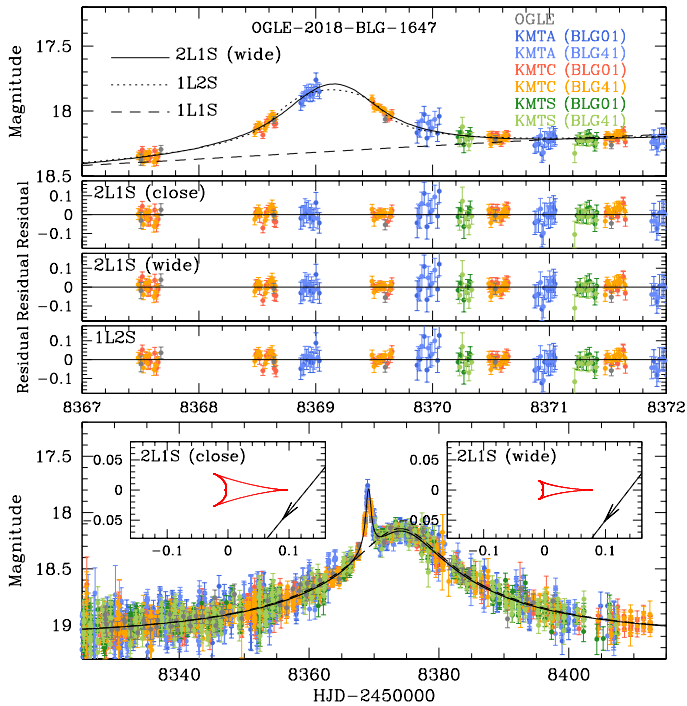}
\caption{Light curve and model for OGLE-2018-BLG-1647.  The anomaly is
a bump centered at 8369.2.  The planetary interpretation is favored
over the binary-source model by
$\Delta\chi^2 =\chi^2({\rm 1L2S})- \chi^2({\rm 2L1S})=28$.
While both close and wide caustic structures are illustrated, the
wide solution is decisively favored by $\Delta\chi^2 =17$.  Nevertheless,
this (albeit broken) degeneracy proved to be the ``Rosetta Stone'' for the
unification of the close/wide and inner/outer degeneracies.  See
Sections~\ref{sec:anal-preamble} and \ref{sec:anal-ob181647}. }
\label{fig:2060lc}
\end{figure}

\begin{table}[t]
\small
\caption{Light Curve Parameters for OGLE-2018-BLG-1647}
\begin{tabular*}{\columnwidth}{@{\extracolsep{\fill}}lcccc}
\hline\hline
\multicolumn{1}{c}{Parameter} &
\multicolumn{1}{c}{Close}     &  
\multicolumn{1}{c}{Wide}      &
\multicolumn{1}{c}{1L2S}      \\
\hline
$\chi^2/$dof                  & $ 9871.51/ 9855$   & $ 9854.69/ 9855$   & $ 9882.81/ 9855$   \\
$t_0 -8300$                   & $ 73.570\pm  0.038$& $ 73.520\pm  0.040$& $ 74.391\pm  0.061$\\
$t_{0,2} -8300$               &                    &                    & $ 69.121\pm  0.006$\\
$u_0\ (10^{-2})$              & $  10.01\pm   0.44$& $  11.00\pm   0.54$& $  11.63\pm   0.63$\\
$u_{0,2}\ (10^{-2})$          &                    &                    & $   0.37\pm   0.31$\\
$t_\e$ (days)                 & $  54.67\pm   1.85$& $  52.31\pm   1.92$& $  57.27\pm   2.21$\\
$s$                           & $  0.794\pm  0.011$& $  1.433\pm  0.014$&                    \\
$q\ (10^{-3})$                & $   9.96\pm   0.65$& $   9.98\pm   0.65$&                    \\
$\langle\log q\rangle$        & $ -2.003\pm  0.028$& $ -2.001\pm  0.028$&                    \\
$\alpha$ (rad)                & $  5.394\pm  0.008$& $  5.365\pm  0.008$&                    \\
$\rho\ (10^{-3})$             & $   3.66\pm   1.31$& $   5.18\pm   1.04$&                    \\
 $\rho_2\ (10^{-3})$          &                    &                    &                    \\
$q_f$                         &                    &                    & $ 0.0362\pm 0.0028$\\
$I_{\rm S}$                   & $  21.01\pm   0.05$& $  20.91\pm   0.06$& $  21.07\pm   0.05$\\
\hline
\end{tabular*}
\label{tab:ob1647parms}
\end{table}

\subsection{{OGLE-2018-BLG-1647}
\label{sec:anal-ob181647}} 

Figure~\ref{fig:2060lc} shows a pronounced bump $\Delta\tau = -0.083$
before the peak.  The grid search returns two local
minima, whose refinements are shown in Table~\ref{tab:ob1647parms}.  
Traditionally, this would be interpreted as the close/wide degeneracy
in which the source passes similar-looking central caustics 
(Figure~\ref{fig:2060lc}), for which
we would expect the geometric mean  to be unity, compared
$\sqrt{s_{\rm close}\times s_{\rm wide}} = 1.07$, for these two 
reported solutions.  On the other hand,
adopting $u_0 = 0.105$, the heuristic
analysis of Section~\ref{sec:anal-preamble} yields 
$\alpha = -52^\circ$ and $s_+^\dagger = 1.07$, i.e., essentially identical
to the geometric mean.  
Hence, this event is much closer mathematically
to the inner/outer degeneracy (derived in the limit of planetary caustics)
than it is to the close/wide degeneracy 
(derived in the limit of central and planetary caustics).

Note that the arithmetic mean of
Equation~(\ref{eqn:heuristic-old}) would yield $(s_+ + s_-)/2=1.11$.
As we discussed in some detail in Section~\ref{sec:anal-preamble},
it was the fact that the geometric mean worked better than the arithmetic
mean that led us to adopt Equation~(\ref{eqn:heuristic}) to unify
the inner/outer and close/wide degeneracies.

Because the wide/inner model is preferred by $\Delta\chi^2=17$, we adopt
it over the close/outer model.  In any case, the two models have
essentially identical mass ratios, $q\simeq 0.010$.  
We also search for 1L2S models, but find that
they are disfavored by $\Delta\chi^2=28$.  See Table~\ref{tab:ob1647parms}.
Hence, they are decisively rejected.

Due to the faintness of the source, we do not attempt a parallax analysis.

OGLE-2018-BLG-1647 is one of three previously known planets that
are listed by \citet{kb190253} as ``in preparation'' but are analyzed here
for the first time.

\begin{figure}[t]
\includegraphics[width=\columnwidth]{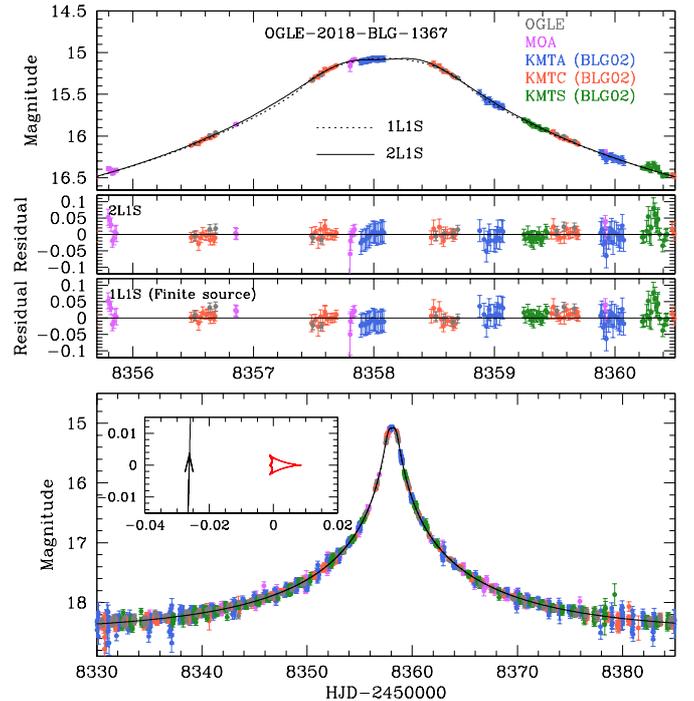}
\caption{Light curve and model for OGLE-2018-BLG-1367.  The anomaly is a
flattening of the peak.  Such flattened peaks can be produced by finite-source
effects in 1L1S events.  However, in this case, the 2L1S interpretation is
favored by $\Delta\chi^2=65$.}
\label{fig:0914lc}
\end{figure}

\subsection{{OGLE-2018-BLG-1367}
\label{sec:anal-ob181367}} 

Figure~\ref{fig:0914lc} shows a flattened, or perhaps slightly
depressed peak.  A natural way to produce a flattened peak is a
1L1S geometry with finite source effects as the lens transits the
face of the source, i.e., so-called finite-source/point-lens (FSPL) events.
We search for such a model, but it produces a relatively poor fit,
$\chi^2({\rm FSPL}) - \chi^2({\rm 2L1S}) = 65$.
In addition, the FSPL fit parameters $(t_\e,\rho)=(22.0\,{\rm day},0.048)$,
would imply an extraordinarily long source self-crossing time
($t_*=1.1\,$day), given that the source is a turnoff star 
(see Section~\ref{sec:cmd-ob181367}).  Hence, the Einstein radius
would be $\theta_\e\simeq 16\,\muas$, while the proper motion
would be an extraordinarily slow $\mu_\rel\simeq 0.27\,\masyr$, with
prior probability
$p=8\times 10^{-5}$.  See Equation~(\ref{eqn:pmulim}), below.
That is, we expect only about one event with such a low proper motion
during the five years of KMT normal observations, and this one event
would have only a few percent chance of giving rise to finite-source effects
(thus enabling its low $\mu_\rel$ to come to our attention).

\begin{table}[t]
\small
\caption{Light Curve Parameters for OGLE-2018-BLG-1367}
\begin{tabular*}{\columnwidth}{@{\extracolsep{\fill}}lcccc}
\hline\hline
\multicolumn{1}{c}{Parameter}    &
\multicolumn{1}{c}{Close}        &
\multicolumn{1}{c}{Wide }        \\
\hline
$\chi^2/$dof                  & $ 7470.68/ 7477$   & $ 7470.73/ 7477$   \\
$t_0 -8350$                   & $ 8.0940\pm 0.0024$& $ 8.0943\pm 0.0024$\\
$u_0\ (10^{-2})$              & $  2.593\pm  0.046$& $  2.598\pm  0.046$\\
$t_\e$ (days)                 & $  22.70\pm   0.18$& $  22.70\pm   0.18$\\
$s$                           & $  0.566\pm  0.044$& $  1.701\pm  0.138$\\
$q\ (10^{-3})$                & $   3.36\pm   0.93$& $   3.22\pm   0.98$\\
$\langle\log q\rangle$        & $  -2.48\pm   0.12$& $  -2.50\pm   0.13$\\
$\alpha$ (rad)                & $  1.604\pm  0.027$& $  1.606\pm  0.031$\\
$\rho\ (10^{-3})$             & $<  16$            & $<  16$            \\
$I_{\rm S}$                   & $  18.89\pm   0.01$& $  18.88\pm   0.01$\\
\hline
\end{tabular*}
\label{tab:ob1367parms}
\end{table}

By contrast, the 2L1S models (Table~\ref{tab:ob1367parms}) fit the data 
quite well and do not
require exceptional physical parameters.  The flattening (or depression)
near the peak is then explained by the source passing roughly perpendicular
to the planet-host axis on the opposite side of the planet, a region that
is characterized by a negative magnification deviation relative to
1L1S.

For perpendicular trajectories, 
$s^\dagger_- = (\sqrt{4+u_0^2}-u_0)/2\rightarrow 0.987$.
Hence, the geometric mean of the two solutions (0.981) is slightly
closer to this value than it is to unity (the close/wide prediction).
This tends to confirm our conjecture that Equation~(\ref{eqn:iocw}) is the
correct generalization of the $s^\dagger$ formalism, even though the event
is qualitatively well described by the ``close/wide'' degeneracy

This is another massive planet, $q\simeq 3.4\times 10^{-3}$, i.e., 3.5
times larger than the Jupiter/Sun ratio.

Because the source is relatively bright and the photometry is good, we attempt 
to measure $\bpi_\e$.  Figure~\ref{fig:parallax} shows the parallax contours
for one of the four solutions, namely the close solution for $u_0>0$.
As in the case of KMT-2018-BLG-2004, the contours are highly elongated
(1-D parallax) with two lobes, indicating that the event is subject
to the jerk-parallax degeneracy.  However, contrary to that case,
the contours do not pass through the origin, but rather cross the $\pi_{\e,N}$
axis at $\pi_{\e,E}\simeq 0.165$, which is 4 times larger than the error.
Hence, this parallax measurement contains significant information.

To extract this information, we follow similar procedures to those
of Section~\ref{sec:anal-kb182004}, but with some difference.
First, contrary to the previous case, there is essentially no bend
between the positive and negative $\pi_{\e,N}$ regimes.  Second,
the contours are essentially identical for 
positive and negative $u_0$.  Third, as mentioned above, the contours
do not pass through the origin.  The first two of these 
imply that there is one regime: 
$(\sigma(\pi_{\e,\parallel}),\psi))=(0.0396,87.30^\circ)$.  To implement the third
within the framework of Equation~(\ref{eqn:pielike}), we rotate the measured
$\pi_{\e,\parallel,0} = 0.165$ to Equatorial coordinates.
\begin{equation}
\bpi_{\e,0}(N,E) = \pi_{\e,\parallel,0}(\cos\psi,\sin\psi) = (0.008,0.165).
\label{eqn:pieeval}
\end{equation}

\begin{figure}[t]
\includegraphics[width=\columnwidth]{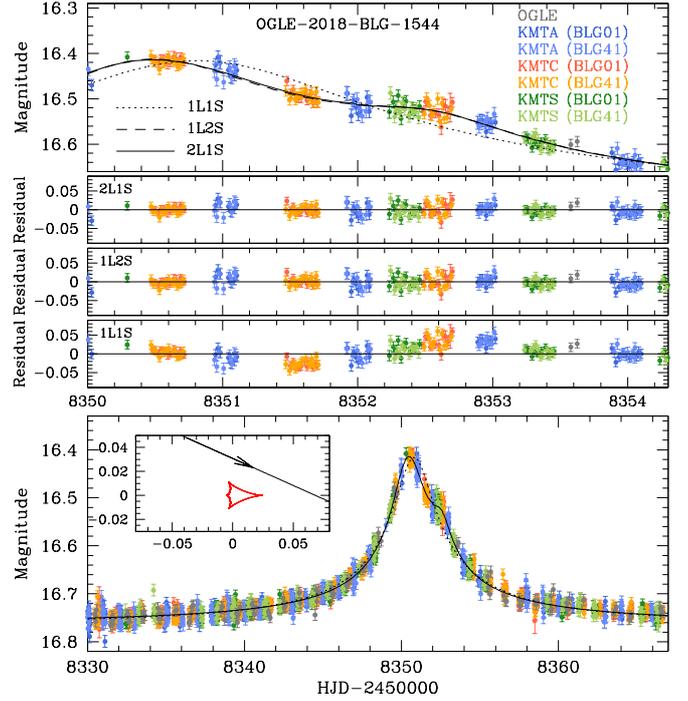}
\caption{Light curve and model for OGLE-2018-BLG-1544, The anomaly is
a long dip near the peak followed by a shorter bump.  The heuristic analysis
is anchored in the latter, which implies a shallow source trajectory
$\alpha=208^\circ$.  The dip is then understood as the lateral passage
of one wall of a central caustic.  See inset.}
\label{fig:0787lc}
\end{figure}

\subsection{{OGLE-2018-BLG-1544}
\label{sec:anal-ob181544}} 

Figure~\ref{fig:0787lc} shows a dip starting near the peak, followed
by a bump centered at $t_{\rm bump}=8352.7$.  If the latter is attributed
to the source crossing the planet-host axis on the planet side, then
the heuristic formalism gives $\alpha=208^\circ$ and $s_+^\dagger = 1.03$.
The angle, in particular, implies that the dip is generated by 
passage along one of the long sides of the central caustic due to
a low-mass (but not necessarily planetary) companion.  In principle,
there might be other geometries.

\begin{table}[t]
\small
\caption{Light Curve Parameters for OGLE-2018-BLG-1544}
\begin{tabular*}{\columnwidth}{@{\extracolsep{\fill}}lcccc}
\hline\hline
\multicolumn{1}{c}{Parameter}    &
\multicolumn{1}{c}{Close}        &
\multicolumn{1}{c}{Wide }        &
\multicolumn{1}{c}{1L2S }        \\
\hline
$\chi^2/$dof                  & $10646.86/10621$   & $10649.97/10621$   & $10652.25/10621$   \\
$t_0 -8350$                   & $  0.879\pm  0.011$& $  0.816\pm  0.010$& $  0.449\pm  0.020$\\
$t_{0,2} -8350$               &                    &                    & $  2.547\pm  0.040$\\
$u_0\ (10^{-2})$              & $   2.85\pm   0.15$& $   2.76\pm   0.15$& $   2.79\pm   0.18$\\
$u_{0,2}\ (10^{-2})$          &                    &                    & $   2.00\pm   0.18$\\
$t_\e$ (days)                 & $  34.51\pm   1.54$& $  35.39\pm   1.66$& $  33.77\pm   1.68$\\
$s$                           & $  0.502\pm  0.020$& $  2.009\pm  0.078$&                    \\
$q\ (10^{-3})$                & $  18.95\pm   2.90$& $  15.66\pm   2.42$&                    \\
$\langle\log q\rangle$        & $ -1.722\pm  0.065$& $ -1.803\pm  0.065$&                    \\
$\alpha$ (rad)                & $  3.562\pm  0.007$& $  3.567\pm  0.006$&                    \\
$\rho\ (10^{-3})$             & $< 13$             & $< 12$             &                    \\
 $\rho_2\ (10^{-3})$          &                    &                    &                    \\
$q_f$                         &                    &                    & $  0.216\pm  0.027$\\
$I_{\rm S}$                   & $  21.48\pm   0.06$& $  21.50\pm   0.06$& $  21.44\pm   0.06$\\
\hline
\end{tabular*}
\label{tab:ob1544parms}
\end{table}

However, the grid search finds only two local minima, which correspond
to the close and wide versions of the one anticipated above, with
$q=0.019$ and $q=0.016$, respectively, the former being favored by
$\Delta\chi^2=3$.  See Table~\ref{tab:ob1544parms}.
Hence, this is another very massive planet
(under the planet definition $q<0.03$).

Due to the faintness of the source, we do not attempt a parallax analysis.

Because this event has a major-image ``bump generating'' caustic
topology, and despite the fact that it does not exhibit the classical 
``isolated bump'' morphology that
would normally induce concerns about a possible binary-source interpretation,
we fit for 1L2S models.  We find that $\Delta\chi^2 = 
\chi^2({\rm 1L2S}) - \chi^2({\rm 2L1S})=5.4$.  See Table~\ref{tab:ob1544parms}.
Hence, while the planetary
interpretation is favored, there is a significant possibility that the
anomaly is actually due to a binary source.

OGLE-2018-BLG-1544 is one of three previously known planets that
are listed by \citet{kb190253} as ``in preparation'' but are analyzed here
for the first time.

\begin{figure}[t]
\includegraphics[width=\columnwidth]{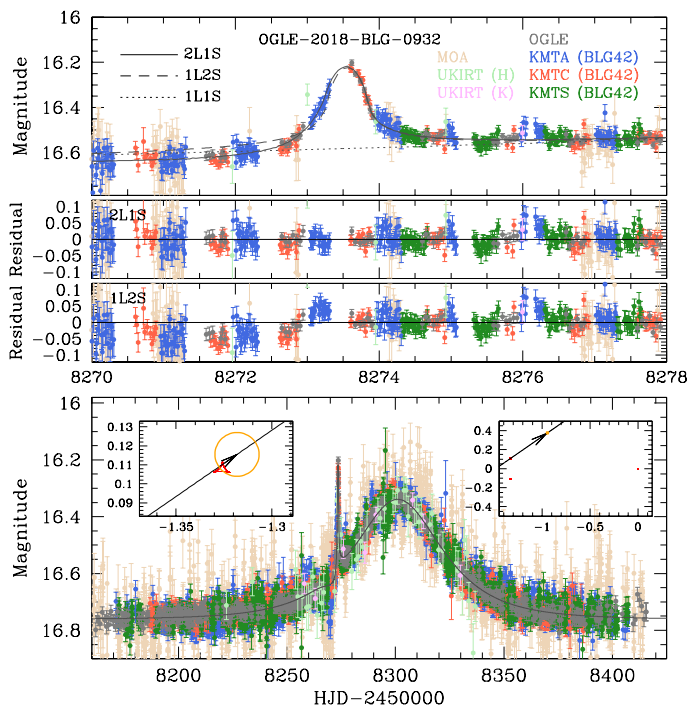}
\caption{Light curve and model for OGLE-2018-BLG-0932.  The anomaly is
a bump centered at 8273.5.  Unlike most smooth, isolated bumps, this
one is due to a source passage over a minor-image caustic, with the
smoothness due to the fact that source is very large compared to the
caustic.  See inset.  Among the 10 events analyzed here, this is the
only one for which the source-size parameter $\rho=\theta_*/\theta_\e$ is
precisely measured.}
\label{fig:0932lc}
\end{figure}

\subsection{{OGLE-2018-BLG-0932}
\label{sec:anal-ob180932}} 

OGLE-2018-BLG-0932 is a good example of a case for which the heuristic
formalism gives relatively imprecise guidance.  The 1L1S approximation has
$(t_0,u_0,t_\e)\simeq (8301.1,0.85,27\,{\rm day})$,  and
$t_{\rm anom}\simeq 8273.5$, i.e., $\tau_{\rm anom}= -1.02$.  These imply
$s^\dagger_\pm = 1.20\pm 0.66$ and $\alpha_+=320^\circ$ (or $\alpha_-=140^\circ$).
The fact that the anomaly is a ``bump'' rather than a ``dip'' leads one
to expect that this is major image perturbation, so $s^\dagger\sim 1.86$,
$\alpha = 320^\circ$.  In fact, however, a full grid search shows that there
is only one solution, for which the bump is due to the source transiting
a triangular caustic from a minor-image perturbation and for which the
heuristic prediction is $s^\dagger_-\sim 0.54$, $\alpha = 140^\circ$.  Comparison
to Table~\ref{tab:ob0932parms} 
shows that $s\simeq s^\dagger_-$, as expected for cases with
no inner/outer degeneracy.  However, $\alpha$ differs from the prediction
by $5^\circ$, which is much larger than any of the other cases examined
here or the 11 cases to which the heuristic analysis was systematically 
applied by \citet{kb190253} and
\citet{kb211391}.  The reason is that the heuristic analysis implicitly
assumes that the anomaly is centered on the planet-host axis.  This
basically holds for major-image planetary perturbations, for dip-like
minor-image planetary perturbations, and even for minor-image caustic
crossings for the cases of very small $q$ (because the caustics are then
very close to the minor-image axis).  However, for the present case, 
$q\sim 10^{-3}$, the caustic is 0.1 Einstein radii from the axis 
(see Figure~\ref{fig:0932lc}, 
i.e., at an angle $\sin^{-1}(0.1/u_{\rm anom})=4^\circ$ relative
to this axis, which accounts for the ``error'' in the heuristic prediction.

\begin{table}[t]
\small
\caption{Light Curve Parameters for OGLE-2018-BLG-0932}
\begin{tabular*}{\columnwidth}{@{\extracolsep{\fill}}lcccc}
\hline\hline
\multicolumn{1}{c}{Parameter}    &
\multicolumn{1}{c}{Value }       \\
\hline
$\chi^2/$dof                  & $ 8893.50/ 8914$   \\
$t_0 -8300$                   & $  1.142\pm  0.047$\\
$u_0\ (10^{-2})$              & $  84.93\pm   0.23$\\
$t_\e$ (days)                 & $  26.88\pm   0.11$\\
$s$                           & $ 0.5355\pm 0.0010$\\
$q\ (10^{-3})$                & $  1.186\pm  0.073$\\
$\langle\log q\rangle$        & $ -2.922\pm  0.026$\\
$\alpha$ (rad)                & $ 2.5339\pm 0.0035$\\
$\rho\ (10^{-3})$             & $  11.66\pm   0.33$\\
$I_{\rm S}$                   & $  16.92\pm   0.01$\\
\hline
\end{tabular*}
\label{tab:ob0932parms}
\end{table}

The results shown in Table~\ref{tab:ob0932parms} 
have blending fixed to zero, specifically
using the baseline source flux as determined by OGLE.  A free fit to
blending gives $f_B/f_{\rm base} = -0.30\pm 0.09$, with an improvement
$\Delta\chi^2=6.7$.  For such a bright source, such large negative blending
cannot be the result of unmodeled fluxes from unresolved stars.  In principle,
it could be a statistical fluctuation (Gaussian probability $p=4\%$), but
is more likely due to low-level systematics or source variability,
or possibly to unmodeled physical effects, such as parallax.

From the present perspective, we simply impose zero blending, while
noting that the parameters do not change much for the negative blending
solutions.  For example, the value of $q$ rises from $1.19\times 10^{-3}$
to $1.26\times 10^{-3}$.  We do not investigate parallax solutions here
because this event has {\it Spitzer} parallax observations under a large
program that was outlined by \citet{yee15}.  These will be analyzed elsewhere.

We searched for 1L2S solutions, but find that these are ruled out
by $\Delta\chi^2=564$.

OGLE-2018-BLG-0932 is one of three previously known planets that
are listed by \citet{kb190253} as ``in preparation'' but are analyzed here
for the first time.

\begin{table*}[t]
\small
\caption{Light Curve Parameters for OGLE-2018-BLG-1212}
\begin{tabular}{lcccc}
\hline\hline
\multicolumn{1}{c}{Parameter}                &
\multicolumn{1}{c}{Close ($\pi_{\e,N}>0$)}   &
\multicolumn{1}{c}{Wide ($\pi_{\e,N}>0$)}    &
\multicolumn{1}{c}{Close ($\pi_{\e,N}<0$)}   &
\multicolumn{1}{c}{Wide ($\pi_{\e,N}<0$)}    \\
\hline   
$\chi^2/$dof                  & $15380.92/15414$   & $15377.95/15414$   & $15392.67/15414$   & $15389.15/15414$   \\
$t_0 -8390$                   & $ 3.7813\pm 0.0015$& $ 3.7607\pm 0.0011$& $ 3.7804\pm 0.0012$& $ 3.7594\pm 0.0012$\\
$u_0\ (10^{-2})$              & $  1.288\pm  0.009$& $  1.373\pm  0.012$& $  1.303\pm  0.010$& $  1.389\pm  0.012$\\
$t_\e$ (days)                 & $  51.19\pm   0.32$& $  51.22\pm   0.32$& $  50.45\pm   0.35$& $  50.47\pm   0.32$\\
$s$                           & $  0.680\pm  0.007$& $  1.451\pm  0.016$& $  0.680\pm  0.007$& $  1.452\pm  0.016$\\
$q\ (10^{-3})$                & $  1.233\pm  0.042$& $  1.234\pm  0.042$& $  1.249\pm  0.044$& $  1.254\pm  0.042$\\
$\langle\log q\rangle$        & $ -2.909\pm  0.015$& $ -2.908\pm  0.015$& $ -2.903\pm  0.015$& $ -2.901\pm  0.014$\\
$\alpha$ (rad)                & $  1.104\pm  0.006$& $  1.105\pm  0.006$& $  1.101\pm  0.006$& $  1.102\pm  0.006$\\
$\rho^a\ (10^{-3})$           &                    &                    &                    &                    \\
$\pi_{\e,N}$                  & $  0.534\pm  0.019$& $  0.534\pm  0.019$& $ -0.406\pm  0.019$& $ -0.408\pm  0.019$\\
$\pi_{\e,E}$                  & $  0.550\pm  0.011$& $  0.549\pm  0.011$& $  0.539\pm  0.011$& $  0.541\pm  0.011$\\
$I_{\rm S}$                   & $  18.60\pm   0.01$& $  18.60\pm   0.01$& $  18.59\pm   0.01$& $  18.59\pm   0.01$\\
\hline
\end{tabular}
\label{tab:ob1212parms}
\end{table*}

\begin{table*}[t]
\small
\caption{Light Curve Parameters for KMT-2018-BLG-2718}
\begin{tabular}{lcccc}
\hline\hline
\multicolumn{1}{c}{Parameter}      &
\multicolumn{1}{c}{Close Plane }   &
\multicolumn{1}{c}{Wide Plane }    &
\multicolumn{1}{c}{Close Binary }  &
\multicolumn{1}{c}{Wide Binary }   \\
\hline   
$\chi^2/$dof                  & $ 6993.06/ 6997$   & $ 6992.66/ 6997$   & $ 7008.48/ 6997$   & $ 7005.38/ 6997$   \\
$t_0 -8350$                   & $   5.32\pm   0.14$& $   5.22\pm   0.14$& $   4.23\pm   0.32$& $   4.24\pm   0.18$\\
$u_0\ (10^{-2})$              & $   4.09\pm   0.75$& $   5.95\pm   1.16$& $   6.26\pm   0.98$& $   2.84\pm   0.54$\\
$t_\e$ (days)                 & $ 230.59\pm  41.76$& $ 161.54\pm  28.82$& $ 182.85\pm  27.00$& $ 361.94\pm  77.15$\\
$s$                           & $  0.688\pm  0.009$& $  1.376\pm  0.025$& $  0.296\pm  0.024$& $  6.334\pm  0.628$\\
$q\ (10^{-3})$                & $  13.74\pm   2.31$& $  19.53\pm   3.24$& $ 696.78\pm 207.49$& $1247.79\pm 796.60$\\
$\langle\log q\rangle$        & $  -1.86\pm   0.07$& $  -1.71\pm   0.07$& $  -0.15\pm   0.12$& $   0.11\pm   0.21$\\
$\alpha$ (rad)                & $  1.693\pm  0.011$& $  1.688\pm  0.011$& $  2.387\pm  0.033$& $  3.938\pm  0.015$\\
$\rho\ (10^{-3})$             & $<  6.8$           & $<  6.8$           & $< 13.0$           & $<  8.7$           \\
$I_{\rm S}$                   & $  23.08\pm   0.20$& $  22.66\pm   0.21$& $  22.71\pm   0.17$& $  23.08\pm   0.26$\\
\hline
\end{tabular}
\label{tab:kb2718parms}
\end{table*}

\begin{figure}[t]
\includegraphics[width=\columnwidth]{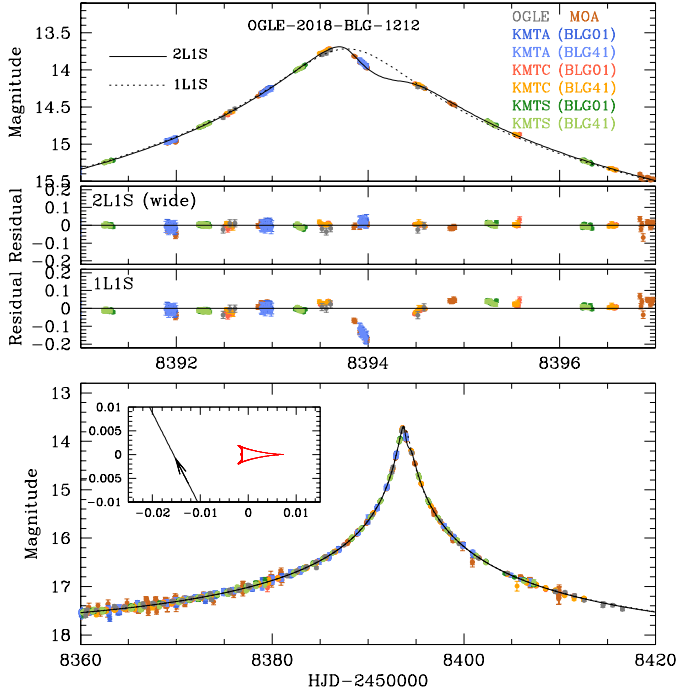}
.\caption{Light curve and model for OGLE-2018-BLG-1212.  The anomaly
is a dip centered at 8394.1, which is traced by both KMTA and MOA data.
The event has a very strong parallax signal and large parallax parameter,
$\pi_\e = 0.767\pm 0.019$, almost certainly implying a nearby lens.
See Sections~\ref{sec:anal-ob181212}, \ref{sec:cmd-ob181212}, and
\ref{sec:phys-ob181212}.}
\label{fig:1212lc}
\end{figure}

\begin{figure}[t]
\includegraphics[width=\columnwidth]{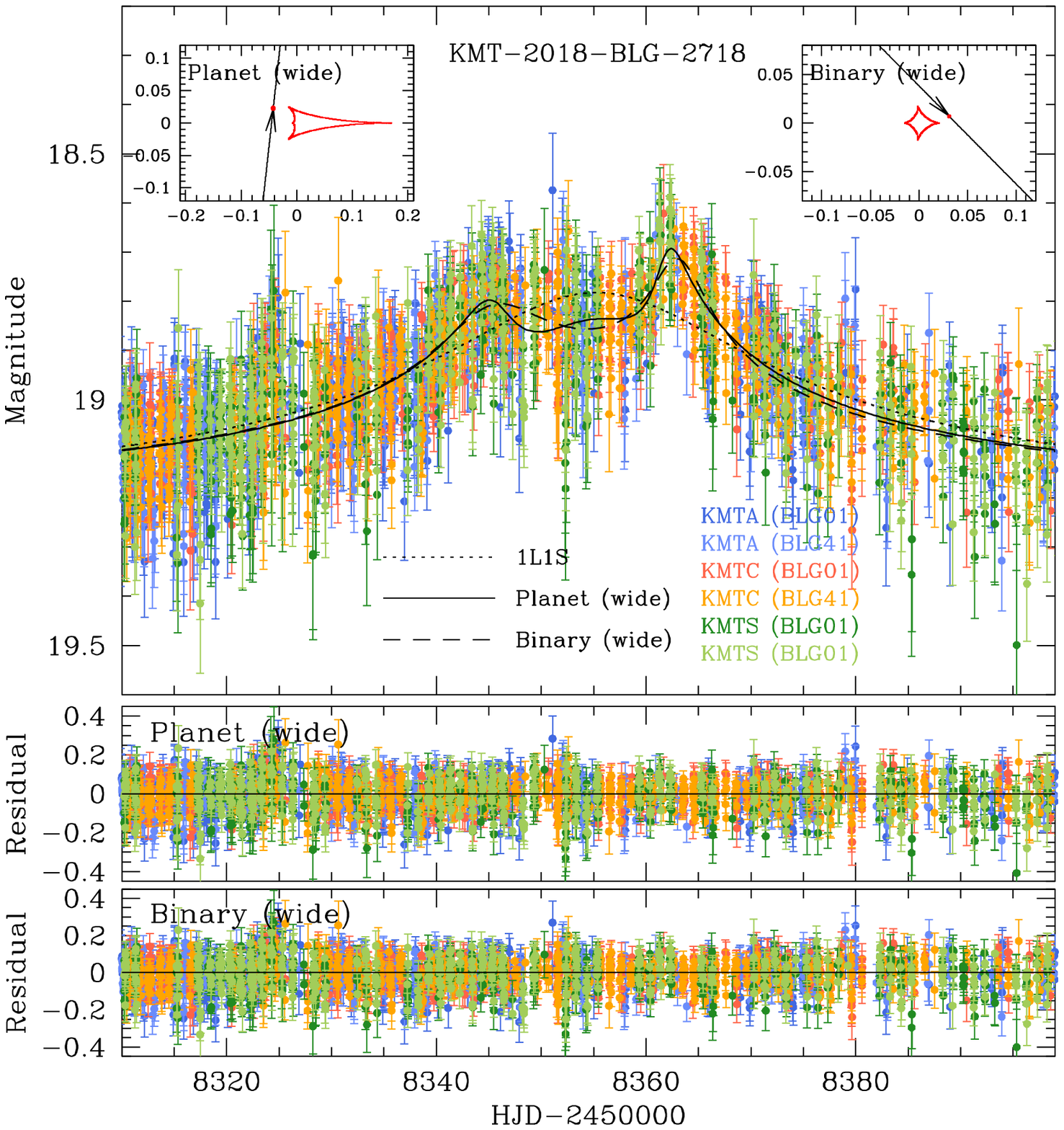}
\caption{Light curve and model for KMT-2018-BLG-2718.  The anomaly
is a dip near the peak, which is flanked by two bumps.
This morphology is the classic signature of the planet/binary degeneracy
identified by \citet{han08}.  See insets.  In this case, the planetary
interpretation is favored by $\Delta\chi^2=12.7$.  The invariant
parameter $t_q = q t_\e=3.16\pm 0.16\,$days would imply non-planetary
mass ratios (by our definition, $q>0.03$), unless $\te\gtrsim 100$~days.
In fact, the fits imply much longer timescales.  
See Table~\ref{tab:kb2718parms}.
}
\label{fig:2718lc}
\end{figure}

\subsection{{OGLE-2018-BLG-1212}
\label{sec:anal-ob181212}}

The light curve for this event shows a strong asymmetry due to parallax,
even when the anomaly is removed.  Hence, contrary to our usual procedures,
we fit for parallax prior to searching for 2L1S solutions.  Both the 1L1S and
2L1S models in Figure~\ref{fig:1212lc} include parallax.  
We can still carry out a
heuristic analysis using the 1L1S parallax-model parameters
$(t_0,u_0,t_\e)=(8393.76,0.014,51\,{\rm day})$, together with the midpoint
and width of the dip: $t_{\rm anom}=8394.1$ and $\Delta t_{\rm dip}=0.75\,$day.
These yield $s_-^\dagger=0.992$, $\alpha=64^\circ$, and $q=7\times 10^{-4}$.
These should be compared with the results from the full grid search
shown in Table~\ref{tab:ob1212parms}, 
i.e., $s^\dagger=\sqrt{s_{\rm close}\times s_{\rm wide}} = 0.993$,
$\alpha= 63^\circ$, and $q=12\times 10^{-4}$.

For the record, we note that in our initial 2L1S fit,
we obtained a very well-localized solution at $\bpi_\e(N,E) = (0.534,0.550)$.
However, we found that the jerk-parallax degeneracy formalism 
(Equations~(7)-(9) of \citealt{mb03037}) predicts another 
solution\footnote{The event peaked almost at quadrature, i.e., $\psi=101^\circ$
in this formalism.  Moreover the projected position of the Sun at this
time is only $0.2^\circ$ from due west, implying that
$\pi_{\e,\perp}\simeq -\pi_{\e,N}$.  Thus, to an excellent approximation,
\citet{mb03037} Equation~(9) becomes $\tilde v_j = (-3/4)v_\oplus/\sin\beta$,
where $\beta = -4.76^\circ$ is the ecliptic latitude.  Hence,
Equation~(7) becomes 
$\pi_{\e,N}^\prime = -\pi_{\e,N} + (4/3)(\au/v_\oplus t_\e))\sin(-\beta)
= -0.534 + 0.126 = -0.408$, very close to the more exact calculation.} 
at $\bpi_\e(N,E) = (-0.404,0.550)$, and numerical investigation then showed
that this was recovered to high precision.  See Table~\ref{tab:ob1212parms}, 
While this second set of solutions is disfavored by $\Delta\chi^2\sim 11$,
we keep track of its potential implications because the $\pi_{\e,\perp}$
($\simeq \pi_{\e,N}$) parameter is among the most sensitive to subtle
systematic errors.

The wide solution is favored by $\Delta\chi^2=3$, which is far below
the level that would be required to distinguish between the two solutions.
However, the parameters (apart from $s$) of the two solutions are essentially
identical.

The very high parallax value $\pi_\e = 0.767$,  
implies a projected velocity 
$|\tilde \bv| \equiv |(\bpi_\e/\pi_\e^2)(\au/t_\e)| = 44\,\kms$ 
in the geocentric frame. 
Noting that Earth's projected velocity at $t_0$ was
$\bv_{\oplus,\perp}(N,E) = (-2.5,-4.6)\,\kms$ and adopting
$\bv_{\odot}(l,b)=(12,7)\,\kms$ for the peculiar velocity of the Sun
relative to the local standard of rest (LSR), 
this implies
$\tilde\bv_\hel (N,E) = (28.2,27.0)\,\kms$ in the Sun frame 
and 
$\tilde\bv_\lsr (l,b) = (50,-2)\,\kms$ in the LSR frame.

This value tends to favor lens distances $D_L\sim 1$--2 kpc.  That is,
ignoring the peculiar motions of the lens relative to the disk and of the source
relative to the bulge, $\tilde \bv_\lsr(l,b)\simeq[(D_S/D_L - 1)^{-1}v_\rot,0]$
for a flat rotation curve with rotation speed $v_\rot = 235\,\kms$.
This would imply 
$D_L\sim (1 + v_\rot/\tilde v_\lsr)^{-1}D_S\rightarrow 1.4\,\kpc$.
Because the lens and source peculiar motions cannot truly be ignored,
and because there is more phase space at larger distances,
this argument is only suggestive.  Nevertheless, we discuss its
potential implications in Sections~\ref{sec:cmd-ob181212} and
\ref{sec:phys-ob181212}.

\begin{table*}[t]
\small
\caption{Light Curve Parameters for KMT-2018-BLG-2164}
\begin{tabular}{lcccc}
\hline\hline
\multicolumn{1}{c}{Parameter}      &
\multicolumn{1}{c}{Close Plane }   &
\multicolumn{1}{c}{Wide Plane }    &
\multicolumn{1}{c}{Close Binary }  &
\multicolumn{1}{c}{Wide Binary }   \\
\hline   
$\chi^2/$dof                  & $ 7655.29/ 7658$   & $ 7655.15/ 7658$   & $ 7659.82/ 7658$   & $ 7660.00/ 7658$   \\
$t_0 -8290$                   & $ 0.9243\pm 0.0070$& $ 0.9237\pm 0.0067$& $ 0.9273\pm 0.0078$& $ 0.9327\pm 0.0072$\\
$u_0\ (10^{-2})$              & $   1.61\pm   0.27$& $   1.62\pm   0.26$& $   1.32\pm   0.18$& $   1.25\pm   0.18$\\
$t_\e$ (days)                 & $  29.19\pm   5.13$& $  29.07\pm   4.47$& $  35.25\pm   4.79$& $  37.06\pm   5.41$\\
$s$                           & $  0.766\pm  0.067$& $  1.302\pm  0.107$& $  0.166\pm  0.045$& $  6.674\pm  1.700$\\
$q\ (10^{-3})$                & $   0.62\pm   0.29$& $   0.66\pm   0.29$& $  86.00\pm  67.27$& $  95.82\pm  73.16$\\
$\langle\log q\rangle$        & $  -3.22\pm   0.20$& $  -3.19\pm   0.19$& $  -1.09\pm   0.31$& $  -1.04\pm   0.32$\\
$\alpha$ (rad)                & $  1.715\pm  0.043$& $  1.715\pm  0.043$& $  5.550\pm  0.037$& $  5.559\pm  0.034$\\
$\rho\ (10^{-3})$             & $<  8.7$           & $<  8.7$           &                    &                    \\
$I_{\rm S}$                   & $  22.74\pm   0.18$& $  22.73\pm   0.17$& $  22.97\pm   0.15$& $  22.97\pm   0.17$\\
\hline
\end{tabular}
\label{tab:kb2164parms}
\end{table*}

\subsection{{KMT-2018-BLG-2718}
\label{sec:anal-kb182718}}

From Figure~\ref{fig:2718lc}, this event does not, at first sight, appear to be
planetary in nature.  The anomaly is a dip near the peak of the event,
which is of very long duration $t_{\rm dip}\sim 20\,$days.  Estimating
$t_\eff\sim 10\,$days and $\Delta t_{\rm anom}\ll t_\eff$ (so $s_-^\dagger \sim 1)$,
we can expect\footnote{The actual value, derived from the MCMCs of 
both close and wide planetary models is $3.16\pm 0.16\,$days.}
 from Equation~(\ref{eqn:qeval2}) that $t_q\simeq 2.5\,$days,
so that this event would only meet our planet definition $q<q_\max=0.03$ provided
that $t_\e \ga t_q/q_\max \sim 83\,$days.  Nevertheless, the
morphology of this very faint ($I_{\rm peak}\sim 18.7$) event does suggest
such a long duration.  This emphasizes the importance of carefully reviewing
all detections of the AnomalyFinder even if they do not look planetary
at first sight.

The grid search indeed returns a wide/close pair of planetary solutions
with $q=0.020$ and $q=0.014$ that are in accord with the above heuristic 
analysis, i.e., with timescales
$t_\e\sim 160\,$days and $230\,$days, respectively.  However, it also returns 
a pair of binary solutions with $q\ga 0.6$.  See Table~\ref{tab:kb2718parms}.
The planetary solutions are favored by $\Delta\chi^2=12.7$.  If the
statistics could be assumed to be Gaussian, then this would decisively
resolve the planet/binary ambiguity.  However, given the quality of the
data and the general inapplicability of Gaussian statistics to microlensing
data, we would rather regard this as ``basically resolved''.

Due to the faintness of the source, we do not attempt a parallax analysis.

\begin{table*}[t]
\small
\caption{Light Curve Parameters for OGLE-2018-BLG-1554}
\begin{tabular}{lccccc}
\hline\hline
\multicolumn{1}{c}{Parameter}      &
\multicolumn{1}{c}{Close Plane }   &
\multicolumn{1}{c}{Wide Plane }    &
\multicolumn{1}{c}{Close Binary }  &
\multicolumn{1}{c}{Wide Binary }   &
\multicolumn{1}{c}{1L2S }          \\
\hline   
$\chi^2/$dof                  & $ 6296.51/ 6309$   & $ 6297.28/ 6309$   & $ 6297.91/ 6309$   & $ 6296.50/ 6309$   & $ 6295.27/ 6309$    \\
$t_0 -8350$                   & $ 4.7965\pm 0.0038$& $ 4.7963\pm 0.0038$& $ 4.7836\pm 0.0036$& $ 4.8102\pm 0.0031$& $ 5.0200\pm 0.0204$ \\
$t_{0,2} -8350$               &                    &                    &                    &                    & $ 4.2855\pm 0.0645$ \\
$u_0\ (10^{-2})$              & $   7.07\pm   0.16$& $   6.93\pm   0.15$& $   6.74\pm   0.11$& $   6.40\pm   0.12$& $   6.22\pm   0.12$ \\
$u_{0,2}\ (10^{-2})$          &                    &                    &                    &                    & $   7.09\pm   0.27$ \\
$t_\e$ (days)                 & $  12.18\pm   0.14$& $  12.37\pm   0.16$& $  12.36\pm   0.13$& $  12.75\pm   0.20$& $  12.36\pm   0.14$ \\
$s$                           & $  0.419\pm  0.046$& $  2.491\pm  0.267$& $  0.288\pm  0.025$& $  3.995\pm  0.508$&                     \\
$q\ (10^{-3})$                & $  21.30\pm   8.63$& $  26.35\pm  10.10$& $  70.55\pm  22.33$& $  83.18\pm  35.63$&                     \\
$\langle\log q\rangle$        & $ -1.675\pm  0.162$& $ -1.584\pm  0.158$& $ -1.151\pm  0.127$& $ -1.073\pm  0.164$&                     \\
$\alpha$ (rad)                & $  1.764\pm  0.023$& $  1.779\pm  0.024$& $  5.881\pm  0.030$& $  5.851\pm  0.020$&                     \\
$\rho\ (10^{-3})$             & $  38.09\pm  10.51$& $  27.28\pm  13.91$&                    &                    &                     \\
$\rho_2\ (10^{-3})$           &                    &                    &                    &                    &                     \\
$q_f$                         &                    &                    &                    &                    & $   0.50\pm   0.12$ \\
$I_{\rm S}$                   & $  19.11\pm   0.02$& $  19.10\pm   0.02$& $  19.12\pm   0.02$& $  19.12\pm   0.02$& $  19.13\pm   0.02$ \\
\hline
\end{tabular}
\label{tab:ob1554parms}
\end{table*}

\begin{figure}[t]
\includegraphics[width=\columnwidth]{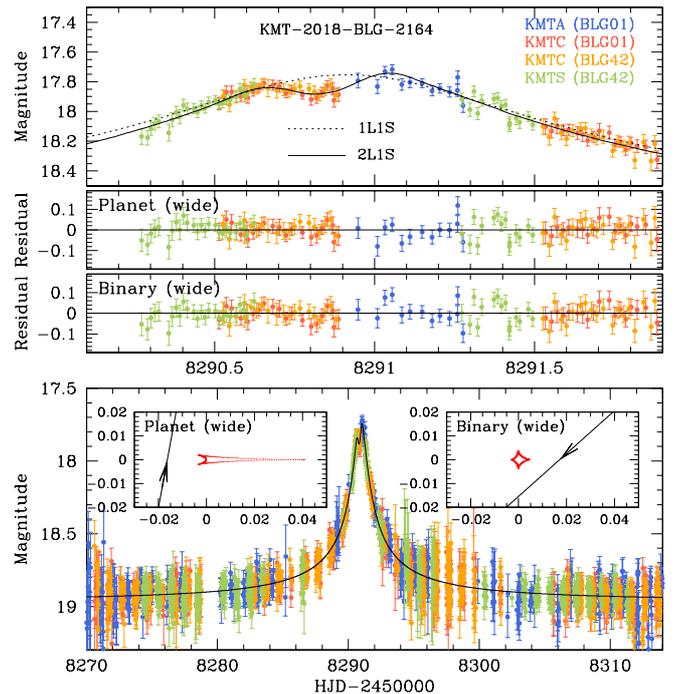}
\caption{Light curve and model for KMT-2018-BLG-2164.  The anomaly is
a dip centered at 8290.8.  Similar to KMT-2018-BLG-2718, this anomaly
is subject to the \citet{han08} planet binary degeneracy  (see insets),
but contrary to that case, the planetary interpretation is not
decisively favored (see Table~\ref{tab:kb2164parms}).  Therefore,
the lens companion cannot be claimed as a planet.
}
\label{fig:2164lc}
\end{figure}

\subsection{{KMT-2018-BLG-2164}
\label{sec:anal-kb182164}}

Figure~\ref{fig:2164lc} shows a dip near the overall peak, flanked
by roughly equal bumps.  In principle, this could be caused
by the source passing roughly perpendicular to the planet-star
axis on the opposite side of the planet, 
similarly to OGLE-2018-BLG-1367.  The grid search indeed returns
a close/wide pair that corresponds to this geometry.  But it
also finds a second pair of minima, in which the source
passes diagonally outside a Chang-Refsdal caustic.  Refinement
of these minima indicate a planet-versus-binary degeneracy,
i.e., $q\sim 0.001$ versus $q\sim 0.15$, which was predicted by 
\citet{han08}.
The planetary solution is favored by $\Delta\chi^2=3.5$, but this
is far below the level what would be required to confidently claim
a planet.  See Table~\ref{tab:kb2164parms}.
This object is presented here because our protocols
demand that we include all companions that are consistent with
being planetary, even if this designation cannot be confirmed.

In this case, the planetary and binary solutions predict similar
source fluxes and there are no proper-motion estimates
(because there is no $\rho$ measurement).  Hence, future adaptic optics (AO) 
observations cannot distinguish between the solutions.  This could only be done
using RV follow-up observations on extremely large telescopes (ELTs), 
or possibly even larger telescopes
that will operate in the more distant future.  Note, however, that even if this
proves to be a planet, the uncertainty in $\log q$ is 0.2 dex, corresponding
to a factor 1.6.  This large uncertainty is related to the fact that
the improvement relative to 1L1S is only $\Delta\chi^2=89$.

Due to the faintness of the source, we do not attempt a parallax analysis.

\begin{figure}[t]
\includegraphics[width=\columnwidth]{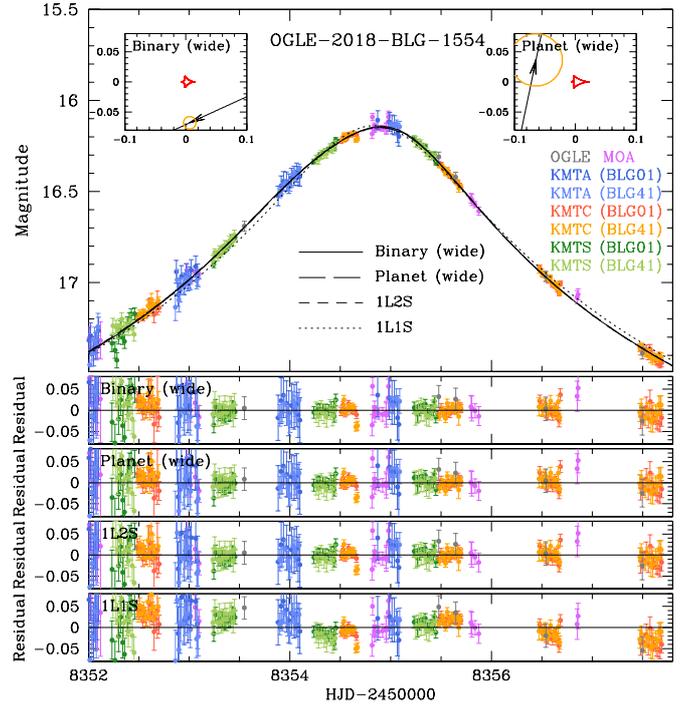}
\caption{Light curve and model for OGLE-2018-BLG-1554.  The anomaly
is characterized by weak deviations both before and after the peak.
Like the previous two events, this one is subject to the \citet{han08}
planet/binary degeneracy (see insets), but even more severely (see
Table~\ref{tab:ob1554parms}).  In addition, there is a severe 1L2S/2L1S
degeneracy.  See Table~\ref{tab:ob1554parms}. Therefore, it is not established
that the lens has a companion, and even if it does, this
companion cannot be claimed as a planet.}
\label{fig:1554lc}
\end{figure}

\subsection{{OGLE-2018-BLG-1554}
\label{sec:anal-ob181554}}

As shown in Figure~\ref{fig:1554lc}, the light curve exhibits a long-term 
deviation over the peak, which is relatively small, but nonetheless we find
to be statistically significant at $\Delta\chi^2 = 413$.  The
grid search returns two pairs of solutions, one being a planetary pair with
$q\sim 0.025$ and the other being a binary pair with $q\sim 0.075$.
In addition to these four solutions, we find a 1L2S solution.  All three
classes have a member that lies within the overall minimum at 
$\Delta\chi^2<1.4$, so all three are ``equally good'' in this sense.
See Table~\ref{tab:ob1554parms}.

Only the planetary solutions have a $\rho$ measurement, $\rho\sim
0.03$, corresponding to $t_*\equiv \rho t_\e \sim 0.4\,$days. In
Section~\ref{sec:cmd-ob181554}, we will show that $\theta_*\simeq
0.93\,\muas$.  Hence, if the planetary solution were correct, then
$\mu_\rel=\theta_*/t_* \sim 0.8\,\masyr$.  As we explain just below
in Section~\ref{sec:cmd}, the fraction of events with such low proper
motions is $p< (\mu_\rel/6.4\,\masyr)^3\simeq 2\times 10^{-3}$.  Thus, we
consider the planetary solution to be extremely unlikely.

In any case, given that the planetary solution cannot (at present) 
be distinguished from
the binary-lens and 1L2S solutions, this event cannot be included in
(present-day) mass-ratio function studies.

For completeness, we remark that if future AO followup observations confirm
the very low $\mu_\rel\la 1\,\masyr$ predicted by the planetary solutions, this
would constitute strong evidence (though not proof) that it was correct.
However, such confirmation would face extreme observational challenges,
even with next-generation 30m class telescopes.

The first point is that if the planetary solution is correct, then 
$\theta_\e \sim 30\,\muas$, and so $\pi_\rel \sim 0.11\,\muas\,(M/M_\odot)^{-1}$.
That is, the lens will be invisibly faint unless the lens and source
are within $D_{LS}\equiv D_S-D_L\simeq D^2_S\pi_\rel/\au\la 100\,\pc$,
which is itself highly improbable.  Moreover, it means that the ``correction''
from the measured geocentric to the relevant heliocentric proper motion,
$\Delta\bmu=\bmu_{\rel,\hel}-\bmu_\rel=\bv_{\oplus\perp}\pi_\rel/\au$ 
will be extremely small.  Here $\bv_{\oplus\perp}(N,E) = (-3.7,+13.7)\,\masyr$
is the projected velocity of Earth at $t_0$.  That is,
$|\Delta\bmu| \sim 0.02(M/0.075\,M_\odot)^{-1}\masyr$, so that 
$\mu_{\rel,\hel}\simeq\mu_\rel = 0.8\masyr$.  Given the faintness of the lens,
this would require waiting of order 3 decades even with ELTs.  Thus,
even in the unlikely case that the planetary solution is correct, the
prospects for confirming it are distant at best.

\begin{table*}[t]
\small
\caption{CMD Parameters}
\begin{tabular}{lcccccccccc}
\hline\hline
\multicolumn{1}{c}{Name}                  & 
\multicolumn{1}{c}{$(V-I)_{\rm S}$}       &
\multicolumn{1}{c}{$(V-I)_{\rm cl}$}      &
\multicolumn{1}{c}{$(V-I)_{\rm S,0}$}     &
\multicolumn{1}{c}{$I_{\rm S}$}           &
\multicolumn{1}{c}{$I_{\rm cl}$}          &
\multicolumn{1}{c}{$I_{\rm cl,0}$}        &
\multicolumn{1}{c}{$I_{\rm S,0}$}         &
\multicolumn{1}{c}{$\theta_*\ (\muas)$}  \\
\hline   
OGLE-2018-BLG-1126& 2.21$\pm$0.07  & 2.48$\pm$0.02 & 0.79$\pm$0.07 & 21.43$\pm$0.07 & 16.40$\pm$0.04 & 14.53 & 19.48$\pm$0.07 & 0.431$\pm$0.042 \\
KMT-2018-BLG-2004 &  1.93$\pm$0.07 & 2.30$\pm$0.02 & 0.69$\pm$0.07 & 19.48$\pm$0.05 & 15.96$\pm$0.04 & 14.46 & 17.98$\pm$0.05 & 0.777$\pm$0.072 \\
OGLE-2018-BLG-1647& 1.97$\pm$0.03  & 2.18$\pm$0.03 & 0.85$\pm$0.04 & 20.87$\pm$0.05 & 15.95$\pm$0.04 & 14.52 & 19.44$\pm$0.07 & 0.471$\pm$0.039 \\
OGLE-2018-BLG-1367& 1.50$\pm$0.04  & 1.87$\pm$0.03 & 0.69$\pm$0.05 & 18.94$\pm$0.02 & 15.30$\pm$0.04 & 14.39 & 18.03$\pm$0.05 & 0.769$\pm$0.059 \\
OGLE-2018-BLG-1544& 2.51$\pm$0.06  & 2.91$\pm$0.02 & 0.66$\pm$0.07 & 21.22$\pm$0.06 & 16.83$\pm$0.04 & 14.44 & 18.83$\pm$0.07 & 0.509$\pm$0.049 \\
OGLE-2018-BLG-0932& N.A.           & N.A.          & 1.05$\pm$0.04 & 16.79$\pm$0.01 & 16.45$\pm$0.04 & 14.41 & 14.75$\pm$0.04 & 5.342$\pm$0.356 \\
OGLE-2018-BLG-1212& 1.45$\pm$0.01  & 1.78$\pm$0.03 & 0.73$\pm$0.02 & 18.52$\pm$0.02 & 15.25$\pm$0.04 & 14.36 & 17.63$\pm$0.05 & 0.956$\pm$0.058 \\
 KMT-2018-BLG-2718&  N.A.          & 2.61$\pm$0.14 & 1.37$\pm$0.14 & 22.60$\pm$0.22 & 15.96$\pm$0.04 & 14.46 & 21.10$\pm$0.22 & 0.358$\pm$0.047 \\
KMT-2018-BLG-2164 &  2.42$\pm$0.05 & 2.30$\pm$0.04 & 1.18$\pm$0.07 & 22.74$\pm$0.15 & 15.98$\pm$0.04 & 14.43 & 21.19$\pm$0.15 & 0.309$\pm$0.048 \\
OGLE-2018-BLG-1554& 2.02$\pm$0.03  & 2.44$\pm$0.03 & 0.64$\pm$0.04 & 19.10$\pm$0.02 & 16.12$\pm$0.04 & 14.49 & 17.47$\pm$0.05 & 0.933$\pm$0.064 \\
\hline
\end{tabular}
\tablefoot{$(V-I)_{\rm cl,0}=1.06$}
\label{tab:cmd}
\end{table*}

\section{{Source Properties}
\label{sec:cmd}}

For a substantial majority of planetary microlensing events that
have been reported in the past, $\rho$ was measured.  Hence,
if the angular source size, $\theta_*$, could be determined, it 
yielded $\theta_\e$ and $\mu_\rel$:
\begin{equation}
\theta_\e = {\theta_*\over \rho}; \qquad
\mu_\rel = {\theta_\e\over t_\e}.
\label{eqn:mu-thetae}
\end{equation}
Then, if $\pi_\e$ could also be measured, one could directly infer
the lens mass and distance via Equation~(\ref{eqn:mpirel}).
However, even if $\pi_\e$ could not be measured, the combination
of $(t_\e,\theta_\e)$ [so, also, $\mu_\rel$] provided more powerful constraints
on the Bayesian mass and distances estimates using Galactic-model priors
than is possible from the $t_\e$ constraint alone.  Moreover, the 
determination of $\mu_\rel$ allows one to accurately estimate how long
one must wait in order to separately resolve the lens and source in
high-resolution follow-up observations using, e.g., AO
on large telescopes or telescopes in space
(e.g., \citealt{ob05169bat,ob05071c,ob05169ben}).

For this reason, virtually all papers on planetary microlensing events
make a serious effort to measure $\theta_*$.  We follow this general
practice here, but we note in advance that, with the exception of two
events (OGLE-2018-BLG-1647 and OGLE-2018-BLG-0932),
the value of doing so is likely to be minimal.
This is because, for all of the other events analyzed here, there are
only weak upper limits on $\rho$, or in some cases no limits at all.

The limit on $\rho$ can be characterized as ``weak'' if it leads to
a ``weak'' lower limit on the proper motion $\mu_{\rm lim} = \theta_*/t_{*,\rm lim}$,
where $t_* \equiv \rho t_\e$ and $t_{*,\lim} \equiv \rho_{\rm lim} t_\e$.
In turn, $\mu_{\rm lim}$ is ``weak'' if it does not exclude a significant
fraction of the parameter space.

We quantify this as follows.  Following the Appendix of \citet{gould21},
we note that for events with bulge lenses and 
bulge sources, the fraction of events
with $\mu_\rel < \mu_{\rm lim}\ll\sigma$ is
\begin{equation}
\eqalign{
p(\mu_\rel < \mu_{\rm lim}) =  & {2\over\sqrt{\pi}}
\int_0^{(\mu_{\rm lim}/2\sigma)^2} z^{1/2}e^{-z} dz \cr
   \rightarrow & {(\mu_{\rm lim}/\sigma)^3\over 6\sqrt{\pi}}
\simeq 4\times 10^{-3}\biggl({\mu_{\rm lim}\over \masyr}\biggr)^3, \cr
}
\label{eqn:pmulim}
\end{equation}
where we have modeled the bulge proper-motion distribution as an isotropic
Gaussian with dispersion $\sigma=2.9\,\masyr$.  One can show that in this
low $\mu_{\rm lim}$ regime, the probability for disk-bulge lensing is
even lower.  Thus, for example, if $\mu_{\rm lim}\la 0.5\,\masyr$ (as in
most of our events) then fewer than
$p\la 10^{-3}$ of simulated events will be eliminated by imposing this
limit, implying negligible impact on the Bayesian estimate.

Nevertheless, while $\theta_*$ is itself of little use in these cases,
the measurements of the source color and magnitude,
which are needed to determine $\theta_*$, can be important for the 
interpretation of future AO observations.  Together, they will enable
prediction of the source flux in the observed band (e.g., $H$ or $J$),
and so allow one to determine which of the two stars is the source,
with the other being the lens, whose properties will be the main
subject of interest.  These observations will, by themselves, yield
$\mu_\rel$ (from the observed separation and elapsed time), and so
$\theta_\e = \mu_\rel t_\e$.  Together with the lens flux, this will
enable good estimates of $M$ and $D_L$.

\begin{figure}
\includegraphics[width=\columnwidth]{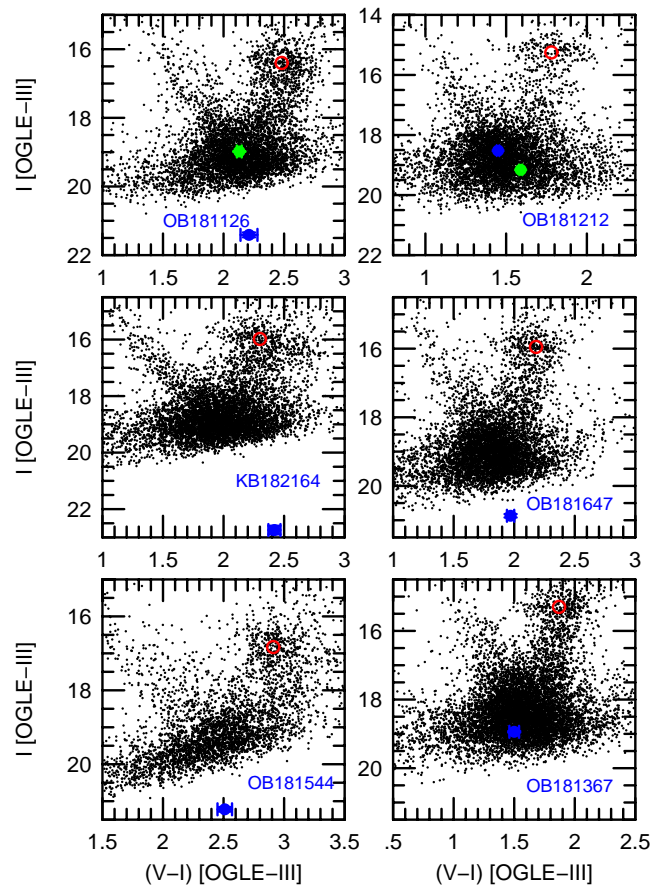}
\caption{CMDs for 6 of the 10 events analyzed in this paper.
The clump centroid is shown in red and the source star is shown
in blue.  Each panel contains an abbreviation of the event name
in blue.  Where relevant, we show the blended light in green.}
\label{fig:allcmd}
\end{figure}

Thus, even though these $\theta_*$ measurements are likely to be of little
use, either now or in the future, they are a small additional step
relative to the actually necessary color and magnitude measurements.
Hence, we report them as well.

\begin{figure}
\includegraphics[width=\columnwidth]{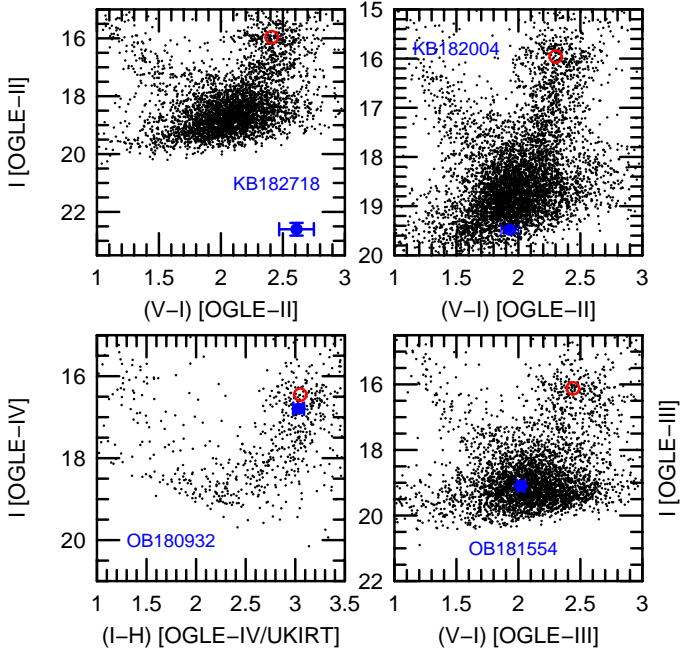}
\caption{Same as Figure~\ref{fig:allcmd} for the remaining 4 (out of 10)
of the events analyzed in this paper.}
\label{fig:allcmd2}
\end{figure}

Our general approach (with a few exceptions that are explicitly noted)
will be to obtain pyDIA \citep{pydia} reductions of KMT data at 
one (or possibly several) observatory/field combinations.  These
yield the microlensing light curve and field-star photometry on the
same system.  We then determine the source color by regression of the
$V$-band light curve on the $I$-band light curve, and the source magnitude
by regression of the $I$-band light curve on the best model.  We then
transform the instrumental KMT photometry to calibrated OGLE photometry,
usually OGLE-III \citep{oiiicat}, but in two cases, OGLE-II 
\citep{oiicat1,oiicat2,oiicat3}.  If there is inadequate $V$-band signal 
in a single observatory/field, we repeat the procedure for several,
check for consistency, and then combine them. In two cases, we are not
able to measure $(V-I)$ from the light curve.  In one of these cases,
we infer the color by combining OGLE-IV $I$-band observations with $H$-band
observations from the UKIRT microlensing project \citep{ukirt17}.  In the
other, we make use of a deep, high-resolution color-magnitude diagram (CMD) 
based on archival
{\it Hubble Space Telescope (HST)} data \citep{holtzman98}.
Figures~\ref{fig:allcmd} and \ref{fig:allcmd2} show 
the resulting CMD for each event, with the position of the source and the
centroid of the red clump indicated in blue and red respectively.
Table~\ref{tab:cmd} lists these values and also shows 
the steps leading to the calculation of $\theta_*$ for each event.

For this, we follow the method of \citet{ob03262}.  We adopt the
intrinsic color of the clump $(V-I)_{0,\rm cl}= 1.06$ from \citet{bensby13}
and its intrinsic magnitude from Table~1 of \citet{nataf13}.
We then obtain 
$[(V-I),I]_{\rm S,0} = [(V-I),I]_{\rm S} + [(V-I),I]_{\rm cl,0} - [(V-I),I]_{\rm cl}$.
We convert from $V/I$ to $V/K$ using the $VIK$ color-color relations of
\citet{bb88} and then derive $\theta_*$ from the color/surface-brightness
relations of \citet{kervella04}.  After propagating errors, we
add 5\% in quadrature to account for errors induced by the overall method.

Where relevant, we report the offset of source from the baseline object.
In all cases, this is found by comparing the difference image near
peak to the baseline object position in the template.

Comments on individual events follow.

\subsection{{OGLE-2018-BLG-1126}
\label{sec:cmd-ob181126}}

The CMD is shown in Figure~\ref{fig:allcmd}.
There are no useful constraints on $\rho$.
We note that the baseline object has $[(V-I),I]_{\rm base}=(2.14,18.70)$,
implying that the blend has $[(V-I),I]_B=(2.13,18.98)$, i.e., 
similar in color but about 9 times brighter than the source.  We
find that it is displaced from the event by 260 mas, meaning that
it is almost certainly unrelated to the event.  Most likely, it is
a bulge sub-giant.  Its brightness and proximity prevent any useful
constraints on the lens flux.  On the positive side, it is unlikely
to interfere with future AO observations.

\subsection{{KMT-2018-BLG-2004}
\label{sec:cmd-kb182004}}

The CMD is shown in Figure~\ref{fig:allcmd2}.
The constraints on $\rho$ have practically no impact.  The baseline object 
($I_{\rm base}=18.88$) is
offset from the source by about 600 mas, meaning that the
blend has $I_B\simeq 19.8$ and is almost certainly unrelated to the event.  
Moreover, the
blend color is very poorly determined.  Hence, we do not display it
in the CMD.  We adopt $I_L>19.6$, which corresponds to
$I_{L,0}>18.1$ for bulge lenses (and other lenses that are behind 
essentially all the dust).  This will have a minor effect.  See
Section~\ref{sec:phys-kb182004}.

The magnitude listed
in Table~\ref{tab:cmd} is for the planetary solution with the lower $\chi^2$,
as will always be the case except when otherwise specified.  In this
case, the other solution would have a larger $\theta_*$ by 1.4\%, i.e.,
a small difference compared to the error bars.

This event is not in the OGLE-III footprint, but fortunately it is in
the OGLE-II footprint \citep{oiicat1,oiicat2,oiicat3}.  
As indicated in Figure~\ref{fig:allcmd2},
we therefore calibrate the photometry using OGLE-II.

\subsection{{OGLE-2018-BLG-1647}
\label{sec:cmd-ob181647}}

The CMD is shown in Figure~\ref{fig:allcmd}.
In this case, there are $\rho$ measurements for both solutions.
Because the wide solution is favored by $\Delta\chi^2=17$, we do not
further consider the close solution.  While the fractional error in $\rho$
is fairly large (20\%), we note that very low values are strongly excluded.
For example, $\rho>0.0023$ at $2.5\,\sigma$, which is very similar to the
naive extrapolation from the $1\,\sigma$ error bar.  This corresponds to
$\theta_\e < 0.20\,\mas$ and
$\mu_\rel  <  1.4\,\masyr$ at the same significance.
Hence, this is likely to be a low-mass lens in the bulge.

OGLE-III photometry, which resolves out a nearby neighbor at about 
600 mas thereby showing a baseline magnitude $I_{\rm base}=19.96$,
implies an estimated blend magnitude $I_B=20.57$.  We set a more
conservative limit on the lens brightness $I_L >20.30$.  Given the
extinction toward this field, $A_I=1.43$, this corresponds
to $I_{L,0} >18.87$ for lenses that that are
behind essentially all the dust.  Hence, given that the $\theta_\e$
measurement already favors a low-mass bulge host, the flux constraint
plays a limited role.  Because we do not have a color determination for
the baseline object (hence, also for the blend), we do not display
it on the CMD.

\subsection{{OGLE-2018-BLG-1367}
\label{sec:cmd-ob181367}}

The CMD is shown in Figure~\ref{fig:allcmd}.
Again, the limit on $\rho$ is very weak, corresponding to
$\theta_\e> 0.048\,\mas$ and $\mu_\rel> 0.77\,\masyr$, which are hardly
constraining.

OGLE-III shows a baseline magnitude $I_{\rm base}=18.57$,
leaving an estimated blend magnitude $I_B=19.92$.  We set a more
conservative limit on the lens brightness $I_L >19.70$, which corresponds
to $I_{L,0}>18.79$ for lenses behind essentially all the dust.  This
is a very similar, mildly constraining limit as in the case of
OGLE-2018-BLG-1647.  Again, we do not display the blend on the CMD
due to poor color determination.

\subsection{{OGLE-2018-BLG-1544}
\label{sec:cmd-ob181544}}

The CMD is shown in Figure~\ref{fig:allcmd}.
The source is blended with a clump giant $[(V-I),I]_{\rm base}=(2.88,16.74)$,
which is separated by 600 mas.  Hence, the blended light cannot be constrained.
Following the logic that was applied to OGLE-2018-BLG-1647,
the limit, $\rho< 0.012$, implies $\mu_\rel >0.45\,\masyr$, which is not useful.

\subsection{{OGLE-2018-BLG-0932}
\label{sec:cmd-ob180932}}

The CMD is shown in Figure~\ref{fig:allcmd2}.
We are not able to accurately measure the $V$-band source flux in spite
of the source being in or near the clump, for two reasons: the source is
heavily reddened and the peak magnification is low ($A_\max=1.47$).  Fortunately,
the event lies in the UKIRT microlensing footprint \citep{ukirt17}, which
allows us to determine the source color on an $[(I-H),I]$ CMD.  To this end,
we match OGLE-IV $I$ and UKIRT $H$ data, which are shown in 
Figure~\ref{fig:allcmd2}.
We find that the source is $\Delta(I-H) = -0.016\pm 0.054$ bluer than
the clump, from which we infer that it is $\Delta(V-I) = -0.01\pm 0.03$,
which is the basis of our color determination in Table~\ref{tab:cmd}.

Note that for this field, $I_{\rm OGLE-III} - I_{\rm OGLE-IV}=0.04$.  We do not
correct for this offset from standard magnitudes in Table~\ref{tab:cmd}.
This makes no difference for our estimate of $\theta_*$, which depends
only on relative photometry.  However, it should be noted in the unlikely event
that there is future, high-precision, $I$-band photometry that could
probe this level of difference.

Of the 10 events analyzed in this paper, OGLE-2018-BLG-0932 is the only
one with a precise $\rho$ measurement and one of only two with any $\rho$ 
measurement.  Combining this with our determination of $\theta_*$, we find,
\begin{equation}
\theta_\e = 0.458\pm 0.033\,\mas\qquad
\mu_\rel = 6.22\pm 0.44\,\masyr .
\label{eqn:thetae-ob180932}
\end{equation}

As discussed in Section~\ref{sec:anal-ob180932}, the blending is consistent
with zero, but is not well measured.  we have set $f_B=0$ in the fit,
but (given that the source is a clump giant), we cannot set any useful
limits on the lens flux.

\subsection{{OGLE-2018-BLG-1212}
\label{sec:cmd-ob181212}}

Before evaluating the CMD information for this event, it is important
to recall that there is a very precise, and fairly large, parallax
measurement $\pi_\e=0.767\pm 0.019$.  As discussed in 
Section~\ref{sec:anal-ob181212}, this result strongly favors (but does
not prove) that the lens is relatively nearby, i.e., only a few
kpc from the Sun.  In light of this, is is notable that the {\it Gaia} 
measurement of the ``baseline object'', 
\begin{equation}
\eqalign{
\pi_{\rm base} = & 3.41\pm 0.93\,  \cr
\bmu_{\rm base}(N,E) = & (-4.73,-7.32)\pm(0.98,1.16)\,\masyr,
}
\label{eqn:gaia1212}
\end{equation}
suggests that the baseline object may be a very
nearby star, or possibly a blend of a nearby object with a more distant
star.  In particular, one scenario is that 
this ``object'' is comprised of a bulge source and a very
nearby disk lens (or a companion to the lens).  
If, for example, the lens contributed half the light (and if the
{\it Gaia} measurement were not itself corrupted, see below), then the 
lens should have $\pi_L\sim 6\,\mas$. In this case, $\pi_\rel\simeq \pi_L$,
in which case the relative proper motion would be
$\mu_\rel = \pi_\rel/\pi_\e t_\e \sim (57\,\masyr)(\pi_\rel/6\,\mas)$.

Such a high lens-source relative proper motion would have two consequences
that are not confirmed.  First, the {\it Gaia} proper motion itself would
be fractionally affected at the same level as the parallax (in this example,
by 50\%), whereas the actual {\it Gaia} proper motion is just $\sim 2\,\sigma$
from the mean of the bulge distribution.  Second, such a high-motion
star would be separately resolved in OGLE-II images (from 1999) and
would be recognizable either as a ``new star'' (at position angle
$\phi\sim 224^\circ$) compared to the
OGLE-IV finding chart (from 2010) or as being displaced from the corresponding
OGLE-IV object in the same direction.  
We find no such high proper motion stars in the OGLE-II images.

Thus, while the large {\it Gaia} parallax may be suggestive of the presence
of a nearby star in the {\it Gaia} 
aperture (whether related to the event of not),
it is difficult to infer anything about the lens from this measurement.
In addition, we note that {\it Gaia} reports a RUWE value of 1.89, probably
indicating some form of contamination of the measurement.

Interestingly, OGLE-2018-BLG-1212 was the subject of a {\it Gaia}
alert\footnote{https://gaia.esac.esa.int/gost/index.jsp}, on 
2018-10-05 08:38:24, as being a transient of unknown origin.
Two of the {\it Gaia} points, at HJD$^\prime=$ 8396.86 and 8396.93, were just
3 days after the anomaly.  However, there are only four significantly
magnified {\it Gaia} points in total.  Hence, these data do not help
constrain the event.

The CMD is shown in Figure~\ref{fig:allcmd}.
There are no useful limits on $\rho$.  The OGLE-III baseline object
has $[(V-I),I]_{\rm base}=(1.50,18.04)$, implying $[(V-I),I]_B=(1.59,19.16)$.
From its position on the CMD, the blend light could very
well be dominated by a companion to the source.   

Of more direct interest, the blend light cannot be dominated by the lens.
For example, given the parallax measurement $\pi_\e\simeq 0.767$,
an $M=0.25\,M_\odot$ lens would lie at $D_L\sim 0.76\,\kpc$, and so
would have roughly $I_L\sim 19$, thus approximately accounting for the
$I_B$ light.  However, after accounting for $E(V-I)_L\sim 0.4$ of reddening,
it would have $(V-I)_L\sim 3.4$, implying $V\sim 22.4$,
which is almost 2 magnitudes redder
than the blend.  On the other hand, if the lens were at $D_L\sim 1.5\,\kpc$
(as crudely estimated in Section~\ref{sec:anal-ob181212} based on
kinematic arguments), then $M\sim 0.11\,M_\odot$.  In this case, the lens
would not contribute significantly to the blended light, thereby avoiding
all photometric constraints.  In principle, the lens could be farther
and so have yet lower mass, but these distances are disfavored by both
the declining mass function and the kinematic arguments.  These will
automatically be taken into account when we carry out a Bayesian analysis
in Section~\ref{sec:phys-ob181212}.  

Thus, in spite of the several intriguing facts about the blend, in the
end, its only implication for the analysis is that it places an
upper limit on the lens light, for which we adopt $I_L>19.0$.  
However, as we discuss in
Section~\ref{sec:phys-ob181212}, even this role has a relatively
modest practical effect.

Finally we note that the proper motion can be expressed
$\mu_\rel = \theta_\e/t_\e = \kappa M \pi_\e/t_\e$, implying,
$\mu_{\rel,\hel} =  45 (M/M_\odot)\masyr$.
Hence, if the lens is luminous $(M\ga 0.075\,M_\odot)$, then its proper
motion is $\ga 3.3\,\masyr$.  Therefore, it will be separated from the
source by at least 40 mas by 2030, a plausible first light for AO on
ELTs.  Note that even if the lens were a white dwarf (WD), it would almost
certainly be visible in AO follow-up.  For example, at $M=0.6\,M_\odot$,
a relatively dim WD with $M_K=14$, would be at $D_L\sim 0.33\,\kpc$ and so
$K\sim 21.6$, which would be visible in ELT observations.  In this case,
the proper motion would be $\mu_{\rel,\hel} = 27\,\masyr$, so that the separation
in 2030 would be $\sim 300\,\mas$.  Hence, a second epoch would be
required for confirmation.  Nevertheless, this does mean that a non-detection
in ELT AO follow-up would imply that the host is a brown dwarf.

\subsection{{KMT-2018-BLG-2718}
\label{sec:cmd-kb182718}}

The CMD is shown in Figure~\ref{fig:allcmd2}.
Due to the small variation in the $V$-band light curve, our standard
procedure for determining the source color yields a very imprecise
result: $(V-I)_{0,S} = 1.54 \pm 0.33$.  We therefore estimate the color
from the $I$-band offset between the source and the clump, which
yields $(V-I)_{0,S} = 1.37 \pm 0.14$, using the Galactic bulge CMD
derived from {\it HST} observation by \citet{holtzman98}.
(As usual, all aspects of this evaluation are based on the lowest-$\chi^2$
solution, i.e., the planetary solution with $s>1$.)

For the four solutions, the limits on $\rho$ shown in 
Table~\ref{tab:kb2718parms}  correspond
to $t_* = (1.6,1.1,2.4,3.1)\,$days.  For the second of these, i.e., 
the best fit, this corresponds to $\mu_\rel>0.12\,\masyr$.  The excluded
region contains a fraction 
$p<(\mu_{\rel,\lim}/2.9\,\masyr)^3/6\sqrt{\pi}=7\times 10^{-6}$.  That is,
this limit is completely unconstraining.  For the other three cases,
the limit is even weaker.

It is unlikely that the ambiguity between planetary and binary
solutions can be decisively resolved until RV observations become
feasible for this very faint host.  Because the planetary solution
is formally favored by $\Delta\chi^2=12.7$, the event can plausibly
be included in mass-ratio function studies.  However, this will
require a specific decision.

\subsection{{KMT-2018-BLG-2164}
\label{sec:cmd-kb182164}}

The CMD is shown in Figure~\ref{fig:allcmd}.
There are no useful constraints on $\rho$.  The OGLE-III baseline object
has $I_{\rm base}=20.54$, yielding $I_B=20.70$.  We adopt $I_L>20.40$, 
corresponding to $I_{L,0}>18.85$ for lenses lying behind essentially all the
dust, which is mildly constraining.  We remind the reader that there
is a factor $\sim 200$ ambiguity in $q$ for the two classes of solutions
that we presented in Section~\ref{sec:anal-kb182164}, which cannot
be resolved except by RV observations in the far future.  Hence, we
believe that this event is unlikely to attract interest for AO follow-up
observations.

\begin{table*}[t]
\small
\caption{Physical properties}
\begin{tabular}{lccccccccccccc}
\hline\hline
\multicolumn{1}{c}{Event}                &
\multicolumn{4}{c}{Physical Parameters}  &
\multicolumn{2}{c}{Relative Weights}     \\
\multicolumn{1}{c}{Models}                                &
\multicolumn{1}{c}{$M_{\rm host}$ $[M_\sun]$}             &
\multicolumn{1}{c}{$M_{\rm planet}$ $[M_{\rm Jup}]$}      &
\multicolumn{1}{c}{$D_{\rm L}$ [kpc]}                     &
\multicolumn{1}{c}{$a_\bot$ [au]}                         &
\multicolumn{1}{c}{Gal.Mod.}                              &
\multicolumn{1}{c}{$\chi^2$}                              \\                      
\hline   
OB181126                                                                                                                                            \\[0.5ex]
Close                 & $0.69^{+0.41}_{-0.38}$    & $0.060^{+0.035}_{-0.033}$ & $5.70^{+1.82}_{-2.42   }$ & $2.96^{+0.95}_{-1.26   }$ &1.00 & 1.00  \\[0.5ex]
Wide                  & $0.69^{+0.41}_{-0.38}$    & $0.043^{+0.025}_{-0.024}$ & $5.69^{+1.82}_{-2.42   }$ & $4.01^{+1.29}_{-1.71   }$ &0.99 & 0.35  \\[0.5ex]
{\bf Adopted}         & $0.69^{+0.41}_{-0.38}$    & $0.056^{+0.033}_{-0.031}$ & $5.70^{+1.82}_{-2.42   }$ & $3.23^{+1.29}_{-1.71   }$ &     &       \\[0.5ex]
\hline                                                                                                                                               
KB182004                                                                                                                                            \\[0.5ex]
$u_0>0$ (inner)       & $0.69^{+0.31}_{-0.31}$    & $0.30^{+0.14}_{-0.14}$    & $6.97^{+1.04}_{-1.53   }$ & $3.78^{+0.56}_{-0.83   }$ &1.00 & 0.34  \\[0.5ex]
$u_0>0$ (outer)       & $0.69^{+0.31}_{-0.31}$    & $0.27^{+0.12}_{-0.12}$    & $6.97^{+1.04}_{-1.53   }$ & $4.95^{+0.74}_{-1.09   }$ &0.99 & 1.00  \\[0.5ex]
$u_0<0$ (inner)       & $0.69^{+0.31}_{-0.32}$    & $0.30^{+0.13}_{-0.14}$    & $6.97^{+1.04}_{-1.52   }$ & $3.77^{+0.56}_{-0.82   }$ &1.00 & 0.34  \\[0.5ex]
$u_0<0$ (outer)       & $0.69^{+0.32}_{-0.31}$    & $0.27^{+0.12}_{-0.12}$    & $6.98^{+1.04}_{-1.53   }$ & $4.90^{+0.73}_{-1.07   }$ &0.99 & 1.00  \\[0.5ex]
{\bf Adopted}         & $0.69^{+0.32}_{-0.31}$    & $0.27^{+0.12}_{-0.12}$    & $6.98^{+1.04}_{-1.53   }$ & $4.62^{+0.80}_{-1.10}$    &     &       \\[0.5ex]
\hline                                                                                                                                              
OB181647              & $0.092^{+0.170}_{-0.053}$ &$0.97^{+1.78}_{-0.56}$     & $7.88^{+1.18}_{-1.00}$    & $1.36^{+0.20}_{-0.17}$    & 1.00 & 1.00 \\[0.5ex]
\hline                                                                       
OB181367                                                                                                                                           \\ [0.5ex]
Close                 & $0.28^{+0.22}_{-0.13}$    & $0.99^{+0.79}_{-0.46}$    & $5.37^{+1.50}_{-1.42   }$ & $1.21^{+0.34}_{-0.32   }$ &1.00 & 1.00 \\ [0.5ex]
Wide                  & $0.28^{+0.22}_{-0.13}$    & $0.95^{+0.75}_{-0.44}$    & $5.37^{+1.50}_{-1.42   }$ & $3.63^{+1.02}_{-0.96   }$ &1.00 & 0.98 \\ [0.5ex]
{\bf Adopted}         & $0.28^{+0.22}_{-0.13}$    & $0.95^{+0.75}_{-0.44}$    & $5.37^{+1.50}_{-1.42   }$ & (bi-modal)                &     &      \\ [0.5ex]
\hline                                                                                                                                               
OB181544                                                                                                                                           \\ [0.5ex]
Close                 & $0.62^{+0.38}_{-0.36}$    & $12.3^{+7.6}_{-7.1}$      & $6.30^{+1.34}_{-2.13   }$ & $1.55^{+0.33}_{-0.53   }$ &1.00 & 1.00 \\ [0.5ex]
Wide                  & $0.62^{+0.38}_{-0.36}$    & $10.2^{+6.3}_{-5.9}$      & $6.30^{+1.34}_{-2.13   }$ & $6.21^{+1.32}_{-2.10   }$ &1.00 & 0.95 \\ [0.5ex]
{\bf Adopted}         & $0.62^{+0.38}_{-0.36}$    & $11.2^{+6.3}_{-5.9}$      & $6.30^{+1.34}_{-2.13   }$ & (bi-modal)                &     &      \\ [0.5ex]
\hline                                                                                                                                             
OB180932              & $0.72^{+0.29}_{-0.25}$    & $0.89^{+0.36}_{-0.31}$    & $6.62^{+0.91}_{-0.86}$    & $1.75^{+0.24}_{-0.23}$    &1.00 & 1.00 \\ [0.5ex]
\hline                                                                                                                                                 
OB181212                                                                                                                                           \\ [0.5ex]
close, $\pi_{\e,N}>0$ & $0.16^{+0.12}_{-0.10}$    & $0.20^{+0.16}_{-0.13}$    & $1.55^{+1.27}_{-0.54 }$   & $0.86^{+0.70}_{-0.30 }$   &0.98 & 0.23 \\ [0.5ex]
wide, $\pi_{\e,N}>0$  & $0.16^{+0.12}_{-0.10}$    & $0.20^{+0.16}_{-0.13}$    & $1.55^{+1.26}_{-0.54 }$   & $1.82^{+1.47}_{-0.63 }$   &1.00 & 1.00 \\ [0.5ex]
close, $\pi_{\e,N}<0$ & $0.14^{+0.10}_{-0.07}$    & $0.19^{+0.13}_{-0.09}$    & $1.93^{+1.27}_{-0.76 }$   & $0.90^{+0.59}_{-0.35 }$   &0.56 & 0.001\\ [0.5ex]
wide, $\pi_{\e,N}<0$  & $0.14^{+0.11}_{-0.07}$    & $0.19^{+0.14}_{-0.09}$    & $1.96^{+1.25}_{-0.76 }$   & $1.87^{+1.19}_{-0.35 }$   &0.50 & 0.004\\ [0.5ex]
{\bf Adopted}         & $0.16^{+0.12}_{-0.10}$    & $0.20^{+0.16}_{-0.13}$    & $1.55^{+1.26}_{-0.54 }$   & $1.68^{+1.47}_{-0.63 }$   &     &      \\ [0.5ex]
\hline                                                                                                                                            
KB182718                                                                                                                                           \\ [0.5ex]
Close                 & $0.85^{+0.63}_{-0.41}$    & $12.2^{+9.1}_{-5.8}$      & $4.29^{+2.62}_{-2.08   }$ & $2.70^{+1.65}_{-1.31   }$ &0.27 & 0.82 \\ [0.5ex]
Wide                  & $0.82^{+0.57}_{-0.39}$    & $16.8^{+11.7}_{-8.0}$     & $4.49^{+2.51}_{-2.15   }$ & $5.32^{+2.97}_{-2.55   }$ &1.00 & 1.00 \\ [0.5ex]
{\bf Adopted}         & $0.82^{+0.57}_{-0.39}$    & $16.0^{+11.7}_{-8.0}$     & $4.49^{+2.51}_{-2.15   }$ & $4.86^{+2.97}_{-2.55   }$ &     &      \\ [0.5ex]
\hline                                                                                                                                             
\end{tabular}                                                                                                                                  
\label{tab:physall}
\end{table*}

\subsection{{OGLE-2018-BLG-1554}
\label{sec:cmd-ob181554}}

The CMD is shown in Figure~\ref{fig:allcmd2}.
As we discussed in Section~\ref{sec:anal-ob181554}, there is a $\rho$
measurement only for the planetary solution.  We argued that its high
value, $\rho\simeq0.03$, rendered the planetary solution highly unlikely.

The OGLE-III baseline object has $[(V-I),I]_{\rm base}=(2.01,18.99)$,
which is very similar to the source values from Table~\ref{tab:cmd},
$[(V-I),I]_S=(2.02,19.10)$, implying that the source is almost unblended.
We adopt an upper limit on lens light $I_L>22.80$, corresponding
to $I_{L,0} > 21.17$ for lenses lying behind most or all of the dust.
This would be a significant constraint.  However, because the event
is not clearly planetary, this constraint has no practical impact.
See Section~\ref{sec:phys-ob181554}.

\section{{Physical Parameters}
\label{sec:phys}}

None of the 10 events reported in this paper have both $\theta_\e$ and
$\pi_\e$ measurements.  Hence, as is customary for a substantial
majority of microlensing planets, we make Bayesian estimates of
the physical parameters of the system by incorporating priors from
a Galactic model.  In the subsections below, we summarize the
constraints that are derived from the light-curve analysis and CMD
analysis, as reported in Sections~\ref{sec:anal} and \ref{sec:cmd}.
Our general approach is to simulate events based on a Galactic model
and then assign each event a weight (possibly zero) depending on
how well it matches these constraints.  For example, if (as is true of
several events), the only constraint is the measurement of the
Einstein timescale $t_\e\pm \sigma(t_\e)$, 
then the weight of the simulated event $i$, with timescale $t_{\e,i}$ is
$w_i = \exp(-\chi^2/2)$ where $\chi^2 = (t_\e - t_{\e,i})^2/[\sigma(t_\e)]^2$.
The Galactic model is summarized in Section~5 of \citet{han-bayesian}.

In Table~\ref{tab:physall}, we present the resulting Bayesian estimates
of the host mass $M_{\rm host}$, the planet mass $M_{\rm planet}$, 
the distance to the lens system $D_L$, and the planet-host projected
separation $a_\perp$.  For the majority
of events, there are two or more competing solutions.  For these cases
we show the results of the Bayesian analysis for each solution separately,
and we then show the ``adopted'' values below these.  For $M_{\rm host}$,
$M_{\rm planet}$, and $D_L$, these are simply the weighted averages of the
separate solutions, where the weights are the product of the two
factors at the right side of each row.  The first factor is simply
the total weight from the Bayesian analysis.  The second is 
$\exp(-\Delta\chi^2/2)$ where $\Delta\chi^2$ is the $\chi^2$ difference
relative to the best solution.  See \citet{kb211391}. For $a_\perp$,
we follow a similar approach provided that either the individual solutions
are strongly overlapping or that one solution is strongly dominant.  
However, if neither condition is met, we enter ``bi-modal'' instead.

We present Bayesian analyses for 8 of the 10 events, but not for
KMT-2018-BLG-2164 and OGLE-2018-BLG-1554.
See Sections~\ref{sec:phys-kb182164} and \ref{sec:phys-ob181554}.
Figures~\ref{fig:bayes_mass} and \ref{fig:bayes_dl} show histograms
for $M_{\rm host}$ and $D_L$ for these 8 events.

\subsection{{OGLE-2018-BLG-1126}
\label{sec:phys-ob181126}}

The only constraint is the measurement of $t_\e$.  As a result
the histograms of host mass and distance are extremely broad.
See Figures~\ref{fig:bayes_mass} and \ref{fig:bayes_dl}.  The planet
has a similarly broad distribution, but is generally in the Neptune-class
range.  We recall from Section~\ref{sec:anal-ob181126} that the planet
is detected by only $\Delta\chi^2=69$.

\subsection{{KMT-2018-BLG-2004}
\label{sec:phys-kb182004}}

This event has three constraints in addition to the $t_\e$ measurement.
First, there is the 1-D parallax measurement, 
$\pi_{\e,\parallel} = 0\pm \sigma(\pi_{\e,\parallel})$, where the error bar
and orientation $\psi$ of the $\pi_{\e,\parallel}$ measurement
take on 4 pairs of values
that depend on the signs of $u_0$ and $\pi_{\e,N}$, as given just
above Equation~(\ref{eqn:bmatrix}).  In addition, there are 
limits on $\rho$ ($< 0.021$ or $<0.024$) and on lens light, $I_L>19.60$.
The 1-D parallax measurement is incorporated via 
Equations~(\ref{eqn:pielike}) and (\ref{eqn:bmatrix})
as described in Section~\ref{sec:anal-kb182004}.
The $\rho$ constraint implies $\theta_\e\ga 35\,\muas$,
corresponding to $\mu_\rel \ga 0.4\,\masyr$, and hence it plays 
virtually  no role.
The main information comes from the $\pi_{\e,\parallel}$ measurement.
Because this measurement is consistent with $\bpi_\e\sim 0$, bulge
lenses are permitted.  Of course, the contours extend up into
the north-east quadrant of the $\bpi_\e$ diagram, which is preferred
by disk lenses, so these are also permitted.  However, because the parallax
constraint has constant width, it is more restrictive of disk lenses
(which have higher $\pi_\e$) than bulge lenses.  Hence, bulge lenses,
which are already favored by higher phase-space density, receive a further
boost.  Within this context the flux constraint plays a modest secondary
role by eliminating some bulge lenses at the very top of the main sequence.
The planet has a Saturn-class mass, and the system is very likely in, or
at least close to, the bulge.

\subsection{{OGLE-2018-BLG-1647}
\label{sec:phys-ob181647}}

\begin{figure}
\includegraphics[width=\columnwidth]{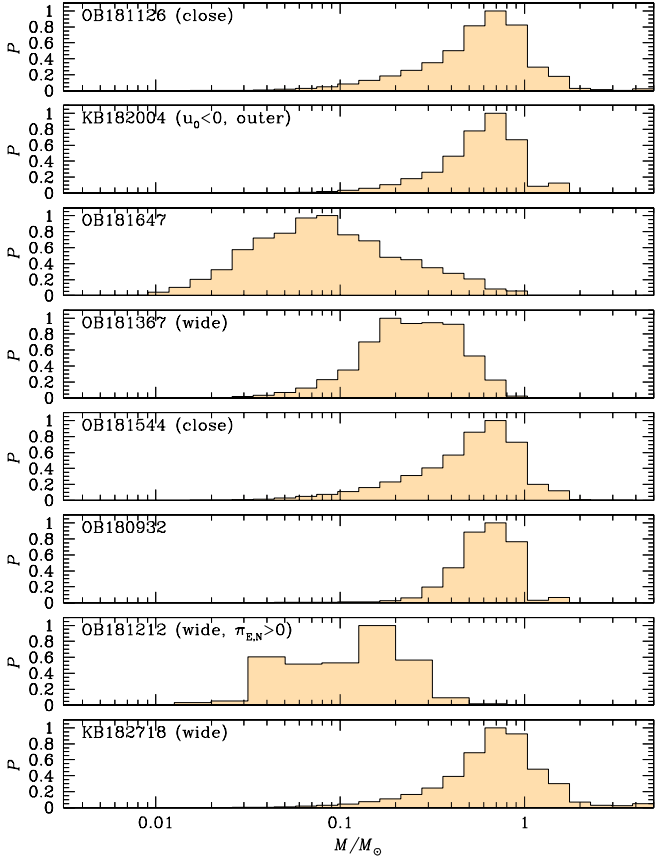}
\caption{Bayesian estimates of $M_{\rm host}$ for the 8 events shown
in Table~\ref{tab:physall}.  Where there are several solutions, we show
the distribution for the one with the lowest $\chi^2$.  However, as
can be assessed from the Table~\ref{tab:physall}, the other solutions
hardly differ.}
\label{fig:bayes_mass}
\end{figure}

The wide solution is favored by $\Delta\chi^2=17$, so we consider
the close/wide degeneracy to be resolved, and so we only show
one solution in Table~\ref{tab:physall}.
Both $t_\e$ and $\rho$ are measured from the light curve, and so
$t_\e$ and $\theta_\e = \theta_*/\rho$ enter as constraints
(Tables~\ref{tab:ob1647parms} and \ref{tab:cmd}).  For the latter
we adopt $\theta_\e = 91\pm 18\,\muas$.  Although the error in this
measurement is large, $\theta_\e$ is nevertheless constrained to
be much smaller than in typical events, which strongly favors 
a low mass $M_{\rm host}\sim 0.1\,M_\odot$ host in or near the Galactic
bulge.  See Figures~\ref{fig:bayes_mass} and \ref{fig:bayes_dl}.
Hence, despite its high mass ratio, $q\simeq 10^{-2}$, the planet
is likely to be of Jovian mass.
We also incorporate the limit on lens light, $I_L>20.30$ from
Section~\ref{sec:cmd-ob181647}.  However, this plays only a small role
because the $\theta_\e$ measurement already heavily disfavors lenses
that are this bright.

\subsection{{OGLE-2018-BLG-1367}
\label{sec:phys-ob181367}}

Like KMT-2018-BLG-2004,
this event has three constraints in addition to the $t_\e$ measurement.
There is 1-D parallax measurement, $\pi_{\e,\parallel} = 0.165\pm 0.040$,
as well as limits on $\rho< 0.016$ and on lens light $I_L>19.70$.
The 1-D parallax measurement is incorporated via 
Equations~(\ref{eqn:pielike}) and (\ref{eqn:bmatrix}) with $\psi=87.30^\circ$,
as described in Section~\ref{sec:anal-ob181367}.
The $\rho$ constraint implies $\theta_\e>48\,\muas$,
corresponding to $\mu_\rel > 0.8\,\masyr$, and hence it plays almost no role.
The main information comes from the $\pi_{\e,\parallel}$ measurement.
First, it implies $\pi_\e\geq \pi_{\e,\parallel}\simeq 0.165$, so if the lens
is in the bulge ($\pi_\rel\la 10\,\muas$), then 
$M=\pi_\rel/\kappa\pi_\e^2\la 0.1\,M_\odot$, which greatly reduces the
phase space accessible to bugle lenses. Second, the smallest values of
$\pi_\e$ are in the north-east quadrant of the $\bpi_\e$ diagram,
which is the preferred location of disk lenses.  Hence, the
lens distance distribution broadly peaks in the disk at $D_L\sim 5\,\kpc$
(i.e., $\pi_\rel\sim 120\,\muas$) and so at masses
$M=\pi_\rel/\kappa\pi_\e^2\la 0.5\,M_\odot$.  The flux constraint therefore
plays a relatively minor role because lenses that would violate it are
already heavily disfavored.  The planet is again Jovian class.

\subsection{{OGLE-2018-BLG-1544}
\label{sec:phys-ob181544}}

Nominally, this event has two constraints, a $t_\e$ measurement and
an upper limit on $\rho$.  However, the latter leads to a very
weak proper-motion constraint $\mu_\rel \ga 0.4\,\masyr$, which therefore
plays virtually no role.  As with OGLE-2018-BLG-1126 (which has only
a $t_\e$ measurement), the posterior Bayesian distributions of mass
and distance are extremely broad.  However, because $t_\e$ is smaller
in the present case by a factor $\sim 0.65$, these distributions are
shifted to somewhat lower mass and distances.  See 
Figures~\ref{fig:bayes_mass} and \ref{fig:bayes_dl}.  Because of the
event's high mass ratio, $q\ga 0.01$, the planet mass estimate is centered
near the planet-BD boundary, but with a wide dispersion.

\subsection{{OGLE-2018-BLG-0932}
\label{sec:phys-ob180932}}

In addition to the $t_\e$ measurement, this event has two constraints,
a measurement of $\rho$ (leading to measurements of 
$\theta_\e=0.458\pm 0033\,\mas$ and $\mu_\rel=6.22\pm 0.44\,\masyr$), 
and a {\it Gaia} measurement of the source
proper motion $\bmu_S(N,E) = (-7.53,-8.81)\pm (0.17,0.26)\,\masyr$.
There are also {\it Spitzer} microlensing data for this event, which
should ultimately yield a $\bpi_\e$ measurement.  However, the analysis
of these data is beyond the scope of the present work and will be
presented elsewhere.

\begin{figure}
\includegraphics[width=\columnwidth]{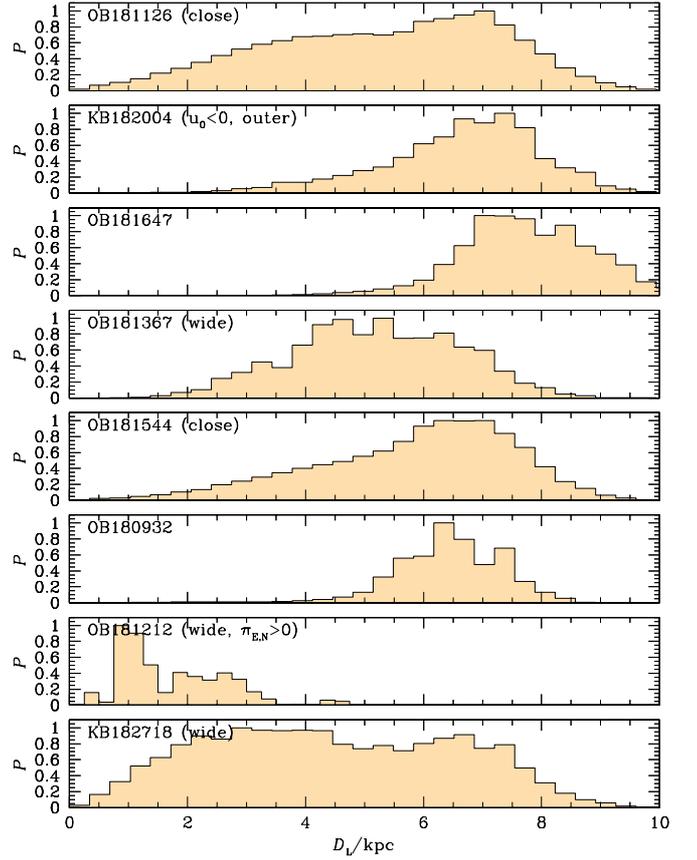}
\caption{Bayesian estimates of $D_L$ for the 8 events shown
in Table~\ref{tab:physall}.  Where there are several solutions, we show
the distribution for the one with the lowest $\chi^2$.  However, as
can be assessed from the Table~\ref{tab:physall}, the other solutions
hardly differ.}
\label{fig:bayes_dl}
\end{figure}

We note that in Galactic coordinates, the source proper motion is 
$\bmu_S(l,b) = (-10.96,+3.78)\,\masyr$, 
which is $\sim 6.2\,\masyr$ from
the bulge mean, i.e., slightly more than $2\,\sigma$
and tending in the direction of anti-rotation.  This means that
a bulge lens would be expected to generate $\mu_\rel\sim 7\,\masyr$
(quite consistent with what is observed), while disk lenses would
be expected to generate $\mu_\rel\sim 11\,\masyr$.  Thus, the {\it Gaia}
measurement increases the likelihood of bulge lenses, which are already
strongly favored by phase-space considerations.  The net result can be judged
from Figures~\ref{fig:bayes_mass} and \ref{fig:bayes_dl}.  The 
planet is in inferred to have Jovian mass.

It will be interesting to compare the host
mass estimate in Table~\ref{tab:physall}
to the results of from the future {\it Spitzer} analysis.
Roughly speaking $M\simeq 0.72\pm 0.27\,M_\odot$ corresponds to
$\pi_\e = \theta_\e/\kappa M = 0.078\pm 0.029$.

\subsection{{OGLE-2018-BLG-1212}
\label{sec:phys-ob181212}}

For this event, there are two constraints in addition to the $t_\e$ measurement.
First, there is a well very well-localized parallax measurement, 
$\pi_\e=0.767\pm 0.019$, whose direction (in the LSR frame) is closely
aligned to Galactic rotation, i.e., $2^\circ\pm 1^\circ$.  Second, there
is a limit on lens flux, $I_L>19.0$.

Table~\ref{tab:physall} shows that the Bayesian mass and distance
estimates are in good accord with the expectations outlined in
Section~\ref{sec:anal-ob181212}, which was based purely on kinematic
arguments.  For comparison, we conducted a separate Bayesian analysis
in which the flux constraint was ignored.  This test showed that inclusion
of the flux constraint drove the distance down from
$1.62\,\kpc$ to $1.55\,\kpc$, i.e., a small effect.

We also show in Table~\ref{tab:physall} the results from the
alternate jerk-parallax solutions.  These have almost no
formal statistical weight, as indicated by the last two columns.
The main point is to show that the principal implications for the
host and planet masses, the system distance, and the projected
separation are not qualitatively different.  In particular, the
planet is a Saturn-class object that is at 1--2 kpc, orbiting a late
M dwarf.  As discussed in Section~\ref{sec:cmd-ob181212}, these predictions
can be tested at first AO light on ELTs.

\subsection{{KMT-2018-BLG-2718}
\label{sec:phys-kb182718}}

Beyond the $t_\e$ measurement that is common to all events, there
is only a weak constraint on the normalized source size, $\rho<0.0068$,
which leads to an exceedingly weak limit on the proper motion, 
$\mu_\rel\ga 0.1\,\masyr$.  Thus, the only real information from the
photometric light curve is that the Einstein timescale is exceptionally long.
Lenses essentially anywhere along the line of sight can generate
such long timescale events by virtue of the rare chance that the
source and lens proper motions are very similar.  At any distance,
large masses $M\propto t_\e^2$ are favored, and these general remarks
are well reflected in the distributions shown in 
Figures~\ref{fig:bayes_mass} and \ref{fig:bayes_dl}.

\subsection{{KMT-2018-BLG-2164}
\label{sec:phys-kb182164}}

KMT-2018-BLG-2164 is neither unambiguously planetary in nature nor
is the planetary interpretation significantly preferred.  That is,
it has only $\Delta\chi^2=4.7$ relative to the binary interpretation.
Even if Gaussian statistics applied, the binary probability would
be $\sim 10\%$.  Therefore, it should not be ``registered as a planet''
in community databases, and we therefore refrain from trying to
characterize it using Bayesian estimates.  It is included in the
present study only for completeness, i.e., to identify all events with
viable planetary solutions, regardless of whether these are unique.

.\subsection{{OGLE-2018-BLG-1554}
\label{sec:phys-ob181554}}

The case for a planetary interpretation for OGLE-2018-BLG-1554 is
even weaker than for KMT-2018-BLG-2164.  First, the 1L2S solution
is slightly preferred by $\chi^2$.  Second, there is a competing
binary solution at $\Delta\chi^2\simeq 0$.  Third, as we remarked in
Section~\ref{sec:cmd-ob181554}, the measured $\theta_\e$ and
$\mu_\rel$ for the planetary (but not binary or 1L2S) solution are
highly unlikely a priori.  Again, this event is only included in
this study for completeness.  We again counsel against its
``registration'' as a planet in community databases, and so
we refrain from a Bayesian characterization.

.\section{{Conclusions}
\label{sec:conclude}}

The goal of this paper was to complete the analysis of all events
from 2018 with viable planetary solutions that were identified by the KMTNet
AnomalyFinder system and that lie in one or more of the 6 KMT prime fields.
Because the main motivation was to prepare a complete sample for statistical
analysis, we pushed the boundaries of this sample beyond what will
ultimately be used in such studies, and we provide sufficient information
to permit future workers to set their own detailed boundaries.  In particular,
we report on all events with viable solutions with mass ratios $q<0.06$,
and we provide detailed analysis of all events that have viable solutions
with $q<0.03$, even for cases that would not normally be published due
to ambiguity with binary-lens ($q>0.03$) and/or binary-source (1L2S) 
solutions.  Indeed, two of the 10 events that we have analyzed are
in one or both of the last two categories and would not normally be
published.  Of the remaining 8 events, two (OGLE-2018-BLG-1544
and KMT-2018-BLG-2004) have
$\Delta\chi^2 = \chi^2({\rm 1L2S}) - \chi^2({\rm 2L1S}) = 5.45$ and 15.1,
respectively, while
another (OGLE-2018-BLG-1126) has almost a factor 2 uncertainty in $q$, which
could lead to their exclusion from future statistical studies.
Of the other 5 planetary events, 2
(OGLE-2018-BLG-0932 and OGLE-2018-BLG-1647) were previously known,
while the remaining 3 (KMT-2018-BLG-2718,
OGLE-2018-BLG-1212, and OGLE-2018-BLG-1367) are new discoveries
by AlertFinder.  These are in addition to the 4 new AlertFinder
discoveries that were previously published (OGLE-2018-BLG-0383,
OGLE-2018-BLG-0506, OGLE-2018-BLG-0516, 
and OGLE-2018-BLG-0977).  There is one additional
AlertFinder recovery, OGLE-2018-BLG-0100, that remains in preparation,
but this has an ambiguous mass ratio $q$ at the factor 100 level.

\begin{table}[t]
\small
\caption{AnomalyFinder Planets in KMT Prime Fields for 2018}
\begin{tabular*}{\columnwidth}{@{\extracolsep{\fill}}lcccl}
\hline\hline
\multicolumn{1}{c}{Event Name} &
\multicolumn{1}{c}{KMT Name}   &
\multicolumn{1}{c}{$\log q$}   &
\multicolumn{1}{c}{$s$}        &
\multicolumn{1}{c}{Reference} \\
\hline
OB180977           & KB180728 & $-4.38$ & 0.88 & \citet{kb190253} \\
OB181185           & KB181024 & $-4.17$ & 0.96 & \citet{ob181185} \\
OB181126$^{a,b}$   & KB182064 & $-4.13$ & 0.85 & This Work \\
OB180506$^{a}$     & KB180835 & $-4.07$ & 0.86 & \citet{kb190253} \\
KB181025$^{b}$     & KB181025 & $-4.03$ & 0.95 & \citet{kb181025} \\
OB180532           & KB181161 & $-4.01$ & 1.01 & \citet{ob180532} \\
OB180516$^{a}$     & KB180808 & $-3.89$ & 1.00 & \citet{kb190253} \\
OB180596           & KB180945 & $-3.74$ & 0.51 & \citet{ob180596} \\
OB180383           & KB180900 & $-3.67$ & 2.45 & \citet{ob180383} \\
KB182004$^{a,c}$   & KB182004 & $-3.43$ & 1.06 & This Work \\
OB181269           & KB182418 & $-3.24$ & 1.12 & \citet{ob181269} \\
OB180932           & KB182087 & $-2.92$ & 0.54 & This Work \\
OB181212$^{a}$     & KB182299 & $-2.91$ & 1.45 & This Work \\
OB180567           & KB180890 & $-2.91$ & 1.81 & \citet{ob180567} \\
KB180748           & KB180748 & $-2.69$ & 0.94 & \citet{kb180748} \\
OB180962           & KB182071 & $-2.62$ & 1.25 & \citet{ob180567} \\
OB181367$^{a}$     & KB180914 & $-2.48$ & 0.57 & This Work \\
OB181011$^{a,d}$   & KB182122 & $-2.02$ & 0.75 & \citet{ob181011} \\
OB181700$^{a,e}$   & KB182330 & $-2.00$ & 1.01 & \citet{ob181700} \\
OB181647           & KB182060 & $-2.00$ & 1.43 & This Work \\
OB181011$^{d}$     & KB182122 & $-1.82$ & 0.58 & \citet{ob181011} \\
OB181544$^{a,c}$   & KB180787 & $-1.72$ & 0.50 & This Work \\
KB182718$^{a}$     & KB182718 & $-1.71$ & 1.38 & This Work \\
\hline
KB182164$^{a,f}$   & KB182164 & $-3.19$ & 1.30 & This Work \\
OB180100$^{g}$     & KB182296 & $-2.58$ & 1.30 & in prep \\
OB181554$^{a,d,f}$ & KB180809 & $-1.67$ & 0.42 & This Work \\ 
\hline
\end{tabular*}
\tablefoot{Event names are abbreviations for, e.g.,
OGLE-2018-BLG-1185 and KMT-2018-BLG-1024.  
a: $s$ degeneracy. 
b: Nearly factor 2 $q$ degeneracy.
c: 1L2S/2L1S degeneracy.
d: Two-planet system.
e: planet in binary system.
f: planet/binary degeneracy.
g: large $q$ degeneracy.}
\label{tab:all2018events}
\end{table}

Table~\ref{tab:all2018events} shows the 26 events with viable planetary
solutions that were recovered or discovered by AnomalyFinder from the
2018 KMT prime-field events.  The 4 previously published discoveries
are from \citet{kb190253} and \citet{ob180383}.  References are given
for the 11 previously published recoveries.  Note that among these,
OGLE-2018-BLG-1700 is marked as a planet in a binary system because
the statistical properties of AnomalyFinder discoveries/recoveries
may differ for such systems.
The 10 entries marked ``This Work'' include 7 discoveries and 3 recoveries,
while one previously known planetary solution remains ``in preparation''.
We consider that the 3 entries below the double line are unlikely
to enter a mass-ratio function analysis, while 5 others 
(OGLE-2018-BLG-1126, KMT-2018-BLG-1025, KMT-2018-BLG-2004, OGLE-2018-BLG-1700, 
and OGLE-2018-BLG-1544)
will require detailed assessment.  Here, we provide only the information
necessary for these assessments but not the assessments themselves.

Through the course of our systematic study of the 10 events
published here, we noticed that the ``$s^\dagger$'' formalism
that was introduced by \citet{kb190253} for heuristic 
analysis should be slightly modified, from using the arithmetic
to the geometric mean of the two solutions.  In this form,
it unifies the so-called close/wide degeneracy of \citet{griest98}
for central and resonant caustics with the so-called inner/outer
degeneracy of \citet{gaudi97} for planetary caustics, a unification
that was previously conjectured by \citet{ob190960}.

\begin{acknowledgements}
This research has made use of the KMTNet system operated by the Korea
Astronomy and Space Science Institute (KASI) and the data were obtained at
three host sites of CTIO in Chile, SAAO in South Africa, and SSO in
Australia.
Work by C.H. was supported by the grants of National Research Foundation 
of Korea (2020R1A4A2002885 and 2019R1A2C2085965).
W.Z. and H.Y. acknowledge support by the National Science Foundation of China (Grant No. 12133005).
The MOA project is supported by JSPS KAKENHI Grant Number JSPS24253004, JSPS26247023, JSPS23340064, JSPS15H00781, JP16H06287, and JP17H02871.
UKIRT is currently owned by the University of Hawaii (UH) and operated by the UH Institute for Astronomy; operations are enabled through the cooperation of the East Asian Observatory. The collection of the 2018 data reported here was supported by NASA grant NNG16PJ32C and JPL proposal \#18-NUP2018-0016.
This paper makes use of data from the UKIRT microlensing surveys \citep{ukirt17} provided by the UKIRT Microlensing Team and services at the NASA Exoplanet Archive, which is operated by the California Institute of Technology, under contract with the National Aeronautics and Space Administration under the Exoplanet Exploration Program.

\end{acknowledgements}

\end{document}